\documentclass[preprint,12pt,3p,sort&compress]{elsarticle}

\usepackage{amssymb}
\usepackage{graphicx}
\usepackage{natbib}
\usepackage{cleveref}
\usepackage{amsmath,amsthm,bm,mathrsfs}
\usepackage{footnote}
\makesavenoteenv{tabular}
\makesavenoteenv{table}

\journal{}

\begin{document}

\begin{frontmatter}

\title{Core regulatory network motif underlies the ocellar complex patterning in \emph{Drosophila melanogaster}}
%\tnotetext[label0]{This is only an example}

\author[label1,label2,label21]{D. Aguilar-Hidalgo\corref{cor1}}
\address[label1]{Max Planck Institute for the Physics of Complex Systems,\\
		N\"{o}thnitzer Strasse 38, 01187 Dresden, Germany.\\}
\address[label2]{Departamento de F\'isica de la Materia Condensada, Universidad de Sevilla.\\ 41012 Sevilla, Spain.}
\address[label21]{CABD (CSIC-UPO-Junta de Andaluc\'ia), 41013 Sevilla, Spain.}
\cortext[cor1]{Corresponding author}
%\fntext[label3]{I also want to inform about\ldots}
%\fntext[label4]{Small city}

\ead{daguilar@pks.mpg.de}
%\ead[url]{author-one-homepage.com}

\author[label2]{M.C. Lemos}
%\address[label5]{Some University}
\ead{lemos@us.es}

\author[label2]{A. C\'ordoba}
\ead{cordoba@us.es}

\begin{abstract}
During organogenesis, developmental programs governed by Gene Regulatory Networks (GRN) define the functionality, size and shape of the different constituents of living organisms. Robustness, thus, is an essential characteristic that GRNs need to fulfill in order to maintain viability and reproducibility in a species. In the present work we analyze the robustness of the patterning for the ocellar complex formation in \emph{Drosophila melanogaster} fly. We have systematically pruned the GRN that drives the development of this visual system to obtain the minimum pathway able to satisfy this pattern. We found that the mechanism underlying the patterning obeys to the dynamics of a 3-nodes network motif with a double negative feedback loop fed by a morphogenetic gradient that triggers the inhibition in a \emph{French flag problem} fashion. A Boolean modeling of the GRN confirms robustness in the patterning mechanism showing the same result for different network complexity levels. Interestingly, the network provides a steady state solution in the interocellar part of the patterning and an oscillatory regime in the ocelli. This theoretical result predicts that the ocellar pattern may underlie oscillatory dynamics in its genetic regulation. 
\end{abstract}


\end{frontmatter}
%\tableofcontents
% \linenumbers

\section{Introduction}
\label{intro}

The study of biological systems from a physical perspective has substantially increased in the last decade as a result of the same increment in the synergy between theoretical and experimental researchers in this field. Physical models help on a better understanding of the principles of biological systems. One of the biological areas that counts with a higher number of theoretical collaborations is developmental biology. This biology field, in the case of multicellular organisms, studies how these organisms develop from a single cell to a complex structure with many differentiated cell types capable of performing distinct functions. Here, the time parameter gives an extra dimension able to show complex emergent properties, and this is indeed very attractive for physicists interested in the study of complex systems. 

Evolutionarily speaking, every developmental process is subjected to mutations that, in some cases, may lead to a modification in the organism that can imply a certain improvement. In many other cases, mutations can lead to a failure and stop of the developmental process. In order to avoid this type of failure, the biological system needs to be as robust as possible to perturbations in its developmental program. Organisms containing a high level of complexity are able to maintain certain modifications between individuals while maintaining robustness in the essential processes. Low complex organisms usually present a clear robustness in all the developmental processes. A noteworthy example is that of the \emph{Caenorhabditis elegans} worm. This worm has exactly 959 somatic cells in adult hermaphrodites and 1031 somatic cells in adult males \cite{Sulston:1977tx}, which is an impressive sign of robustness in its growth process. 

In the present work we are focusing in another organism, \emph{Drosophila melanogaster} (\emph{D. mel.}), also known as the \emph{fruit fly}. In \emph{D. mel.} our interest relies on the robustness of a specific tissue patterning, the ocellar complex. The ocellar complex is part of the visual system of many insects. In this fly, it is formed by three ocelli (or simple eyes), situated at the vertices of a triangular patch of cuticle, and their interocellar region, placed in the dorsal head. This singular patterning is actually formed by the fusion of the dorsal-anterior domains of the two eye discs \cite{Dessaud:2010km}, so it can indeed be studied as the two ocelli pattern of one of the two eye imaginal discs. A one-dimensional description is then proposed where two ocelli regions are separated by an interocellar cuticle. A physical model based on reaction-diffusion equations has already been proposed in \cite{AguilarHidalgo:2013bw} to fulfill, not only the tissue patterning for the Drosophila ocellar complex, but also the correct experimentally tested behavior of the gene regulatory network (GRN) that drives its formation.

This specific pattern follows the \emph{French flag} problem paradigm, where different developmental programs are run depending on the concentration of an extracellular diffusive molecule (morphogen) \cite{Kondo:2010bx}.

Following \citet{Kang:2012wr}, in order to test robustness in a physical model we can identify three classes of perturbations that lead to three types of robustness:

\begin{enumerate}
\item Robustness with respect to changes in the model parameters.
\item Robustness with respect to transient changes.
\item Robustness with respect to changes in the structure of the equation itself.
\end{enumerate}

Types 1 and 2 have been tested in \cite{AguilarHidalgo:2013bw}. The equations there described provide the same pattern when the system is subject to noise perturbation and changes in the initial conditions. Moreover, an extensive parameter sensitivity analysis was performed, where perturbations in the parameter values retrieved the same patterning with changes in size of its components \cite{Kang:2012wr}, what suggests that the same GRN determines a certain phenotypic variation.

The third type of perturbations considers structural changes in the equations describing a physical model. In a networked system, an exercise to test this type of robustness is to modify the wiring of the network by adding/removing links like in \cite{Li:2004el}. This, in a GRN modifies/removes regulatory interactions. As mentioned above, many regulatory interactions related to the ocellar complex developmental GRN were tested in \cite{AguilarHidalgo:2013bw}, and agreed with the model. However, there is an important issue related to structural robustness of the system remaining untested. This is the model description level \cite{Schlitt:2005ka}. As the patterning retrieved by this model is robust facing parameter values modification, noise, changes in the initial conditions and mutations, we are interested to know the reason for such robustness in the patterning retrieval.

In the present work we focus in perturbations on the structure of the equations with respect to the description level of the physical model. The model presented in \cite{AguilarHidalgo:2013bw} describes the behavior of the GRN for the development of the ocellar complex in a high level of detail, including both genes and products representations, complex ligand-receptor dynamics and a nonlinear formulation. Here, we have reduced the description level of the model while maintaining the biological sense according to the results in \cite{AguilarHidalgo:2013bw}, and thus, we have tried to find out the minimum network expression capable of showing the correct pattern. To do this, we have systematically removed nodes and links and, then, changed the formulation from a non-linear to a linear paradigm. 

The patterning seems at this point to be independent on the description level and formulation complexity. Both the patterning and the individual behaviors of each GRN component is correctly fulfilled. An observation on the resulting GRN leads to the extraction from this GRN of a core formed by a three-nodes network motif that satisfies the correct patterning by itself. This core network is biologically very relevant as it defines the main morphogen-signaling pathway in the ocellar development with the addition of its main network repressor in a double negative feedback loop. A further analysis shows that even the reduction of this core network to a simple activator-repressor feedback loop retrieves the desired patterning. All the different models described are then simulated using Boolean networks. This changes the description level of the dynamical system to its simplest representation \cite{Schlitt:2005ka}. This exercise confirms that this patterning is independent on the description level but dependent on specific regulatory topology of the activator-repressor motif.

\section{Results}
\subsection{Description of the system: the ocellar complex pattern}\label{sec:gen_descr}

During fly development, the ocellar complex is formed describing a triangular patch with a simple eye on each vertex: one anterior ocellus (AO) and two posterior ocelli (PO) (Fig. \ref{fig:1}a). It must be clarified that the fly develops two eye discs, and each eye disc forms two ocelli, one PO and one AO. In a late larval stage the two anterior ocelli are fused into one and only anterior ocellus. In order to describe the ocellar complex conformation in a simplified way, it can be reduced to a one-dimensional description by analyzing one row of cells situated in a cross line over the two ocelli of one of the two eye discs. This conformation shows a 1D field compound of five blocks of three different types of tissues. The external tissue will be referred hereinafter as periocellar region (POC). The following inner type of tissue corresponds to the region where the ocelli will be placed; this tissue will be referred as ocellus (OC). The tissue between the two ocelli will be named interocellar cuticle (IOC). The distribution of these types of tissue in the 1D field will be POC-OC-IOC-OC-POC (Fig. \ref{fig:1}b).

As mentioned above, the ocellar complex pattern in \emph{D. mel.} is dependent on a morphogen profile. One of the most studied morphogens is the molecule Hedgehog (Hh), which has already been considered to successfully describe intercellular communication in developmental processes \cite{Nahmad:2009cp}. This morphogen is known to be essential for the formation of the ocellar complex in \emph{D. mel.} \cite{Royet:1996vg}. Here, Hh is expressed in a central stripe of cells in the target tissue diffusing in the anterior-posterior axis. The readout of this system is the retinal determination (RD) genes group. These genes, \emph{eyes absent} (\emph{eya}) and \emph{sine oculis} (\emph{so}), are required for the formation of the ocelli \cite{Bonini:1993ud,Cheyette:1994uf,Serikaku:1994up,Bonini:1998kk,Blanco:2009bg,Blanco:2010fd,Brockmann:2010kg}. Experimental work in Drosophila and in vertebrates indicates that the Hh signaling pathway is subject to extensive feedback among elements of the pathway, most notably that of the Hh receptor \emph{patched} (\emph{ptc}) \cite{Chen:1996wp,Marigo:1996vp}. The gene \emph{engrailed} (\emph{en}) is activated by the Hh signaling pathway and it self-regulates only in the cells where its protein (En) concentration is high. This En high concentration region overlaps with the Hh expression cells though the En domain can vary depending on external conditions and variations in the GRN \cite{AguilarHidalgo:2013bw}. En, then, attenuates the Hh signaling pathway to establish a region with no RD expression (IOC) and thus, the ocellar pattern.

An extensive experimental and theoretical work in the ocellar complex shows robustness in the patterning against noise, changes in the initial conditions and parameter values \cite{AguilarHidalgo:2013bw}. There, some structural modifications were also tested by removing links in the regulatory network. As example, the ocellar pattern is maintained with enlarged ocelli if the repression between Homothorax (Hth) and $eya$ is removed; on the contrary, it is noteworthy to mention an example by which the ocellar pattern is not fulfilled, this takes place when removing the link between Delta (Dl) and En; here the anterior and posterior ocelli fuse and just one big ocellus is developed. These structural modifications were experimentally validated and are also considered in the present work.  The proposed model in \cite{AguilarHidalgo:2013bw} for the GRN in (Fig. \ref{fig:1}c) is implemented using a non-linear formulation and is given by the following equations:

\begin{equation}\label{eq:Hh0}
\frac{\partial Hh}{\partial \tau }=D\frac{\partial^2 Hh}{\partial x^2 }+\delta(x)\alpha_{hh}-\gamma_{Ptc-Hh}Ptc\cdot Hh-\beta_{Hh}Hh
\end{equation}

\begin{equation}\label{eq:ptc0}
\frac{\partial ptc}{\partial \tau }=\kappa_0\,\beta_{ptc}\left(\left(\alpha_{ptc}+\kappa_{Ciptc}\,\phi(CiA\,\psi(CiR,k_{CiR},n_{CiR}),k_{CiA},n_{CiA})\right)\,\psi(En,k_{En-ptc},n_{En-ptc})-ptc\right)
\end{equation}

\begin{equation}\label{eq:pPtc0}
\frac{\partial Ptc}{\partial \tau }=\theta_{ptc}\,ptc-\gamma_{Ptc-Hh}Ptc\cdot Hh-\beta_{Ptc}Ptc
\end{equation}

\begin{equation}\label{eq:PtcHh0}
\frac{\partial PtcHh}{\partial \tau }=\gamma_{Ptc-Hh}Ptc\cdot Hh-\beta_{PtcHh}PtcHh,
\end{equation}

\begin{equation}\label{eq:ci0}
\frac{\partial ci}{\partial \tau }=\kappa_0\,\beta_{ci}\left(\alpha_{ci}\,\phi(\psi(En,k_{En-ci},n_{En-ci}),k_{ci},n_{ci})-ci\right)
\end{equation}

\begin{equation}\label{eq:CiA0}
\frac{\partial CiA}{\partial \tau }=\kappa_0\,\beta_{CiA}\left(\theta_{ci}\,ci-CiA\right)-\kappa_{Ci}\,\phi(CiA\,\psi(\frac{PtcHh}{Ptc},k_{PH},n_{PH}),k_{CiA},n_{CiA})
\end{equation}

\begin{equation}\label{eq:CiR0}
\frac{\partial CiR}{\partial \tau }=\kappa_{Ci}\,\phi(CiA\,\psi(\frac{PtcHh}{Ptc},k_{PH},n_{PH}),k_{CiA},n_{CiA})-\kappa_0\,\beta_{CiR}\,CiR
\end{equation}

\begin{equation}\label{eq:en0}
\frac{\partial en}{\partial \tau }=\kappa_0\,\beta_{en}\left(\alpha_{en}+\phi(CiA\psi(CiR,k_{CiR},n_{CiR}),k_{CiA},n_{CiA})+\kappa_{En}\,\phi(En,k_{DlEn},n_{En})-en\right)
\end{equation}

\begin{equation}\label{eq:En0}
\frac{\partial En}{\partial \tau }=\theta_{en}\,en-\beta_{En}En
\end{equation}

\begin{equation}\label{eq:eya0}
\frac{\partial eya}{\partial \tau }=\kappa_0\,\beta_{eya}\left(\alpha_{eya}+\alpha_{Toy}\,\phi(CiA\psi(Hth,k_{Hth-eya},n_{Hth}),k_{CiA-eya},n_{CiA})+\phi(Eya,k_{Eya},n_{Eya})-eya\right)
\end{equation}

\begin{equation}\label{eq:Eya0}
\frac{\partial Eya}{\partial \tau }=\theta_{eya}\,eya-\beta_{Eya}Eya
\end{equation}

\begin{equation}\label{eq:hth0}
\frac{\partial hth}{\partial \tau }=\kappa_0\,\beta_{hth}\left(\alpha_{hth}+\alpha_{Wg}\,\phi(\psi(Eya,k_{Eya-hth},n_{Eya}),k_{Wg},n_{Wg})-hth\right)
\end{equation}

\begin{equation}\label{eq:Hth0}
\frac{\partial Hth}{\partial \tau }=\theta_{hth}\,hth-\beta_{Hth}Hth
\end{equation}

\begin{equation} \label{eq:deltax0}
\delta(x) = \left\{
	       \begin{array}{ll}
		 1      & \textmd{if\ } x \in \textmd{\emph{hh}-expressing cells} \\
		 0      & \textmd{if\ } x \notin \textmd{\emph{hh}-expressing cells}\\
	       \end{array}
	     \right.
\end{equation}

where $\phi(X,k,n) = \frac{X^n}{k^n+X^n}$ and $\psi(X,k,n) = 1-\phi(X,k,n)$, with X any system variable. The model contains different parameter types: $\alpha_X$ for the basal transcription rates, $\beta_X$ for
the degradation rates, $k_X$ for the Hill equation transcriptional regulators, $n_X$ for the Hill
coefficients, $\theta_X$ for the translation rates; $\gamma_X$ for protein complex formation. The non-dimensional
parameters $\kappa_0$, $\kappa_{Ci}$, $\kappa_{En}$ and $\alpha_{ci}$ are used for changing the scale of different terms and $D$ is the diffusion coefficient. Subscript X-Y, with X and Y system variables,
indicates a regulation from X to Y.

\subsection{Robustness in the patterning circuit against changes in the structure of the system}

Here we propose a series of modifications in the structure of the system developed in \cite{AguilarHidalgo:2013bw} in order to test robustness in a different way as done there and to try to infer the basic mechanism responsible of the patterning. These changes include pruning of the GRN by eliminating certain nodes and links. These modifications assume changes in the behavioral paradigm of some regulatory elements that, while maintaining the overall concept, they simplify the corresponding mechanisms. Additionally, changes in the formulation paradigms are also being considered.

\subsubsection{Reduction of the GRN leads to proper patterns}

\begin{figure}
\begin{center}
\includegraphics[width=\textwidth]{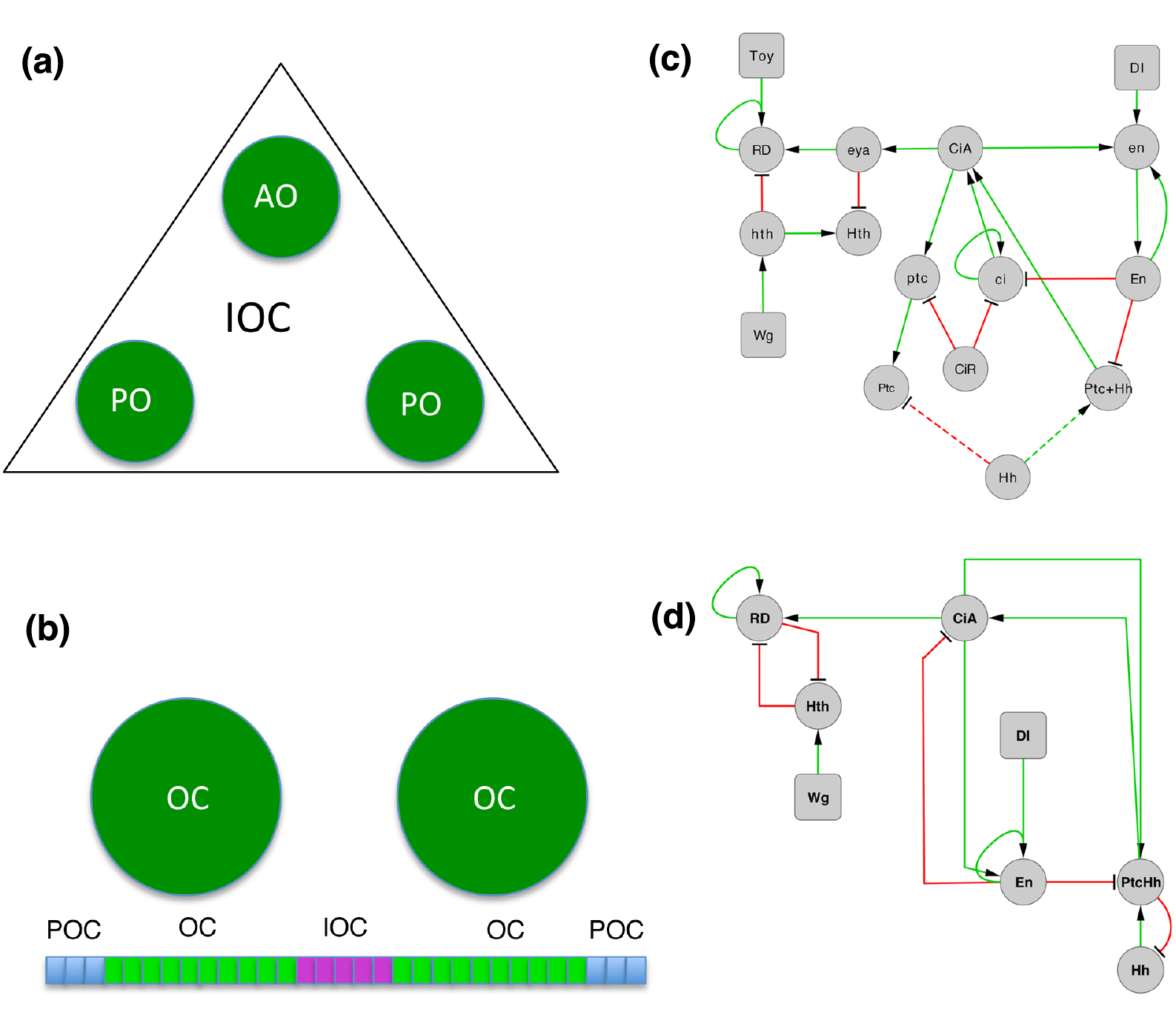}
\end{center}
\caption{(a) Ocellar complex scheme showing the two posterior ocelli (PO) and the anterior ocellus (AO). The region among the ocelli is called the interocellar region. The black triangle represents the triangular cuticle that embraces the ocellar complex. (b) Scheme of the one-dimensional simplification of the ocellar complex. The two ocelli (AO and PO) are considered identical (OC) so the model is symmetric. The one dimension ocellar complex is formed by three different types of tissues represented by POC, or periocellar region (in blue), OC, or ocellar region (in green) and IOC, or interocellar region (in purple). (c-d) Network diagrams of the ocellar model representing the system's variables (circle nodes) and constant inputs (square nodes) linked by positive (green arrows) and negative (red arrows) interactions. (c) GRN in \cite{AguilarHidalgo:2013bw}. Dashed lines from Hh nodes implies that Hh contributes to the formation of the PtcHh complex by repressing the number of free receptor Ptc. (d) Simplified GRN.}
\label{fig:1}
\end{figure}

In order to find the basic mechanism to satisfy the ocellar pattern it is possible to systematically reduce the GRN in \cite{AguilarHidalgo:2013bw} (Fig. \ref{fig:1}c) by removing nodes and links as the model is simplified and the system of equations is reduced.

To simplify this system we firstly focus on the receptor dynamics. This previous model implemented a detailed and complex receptor dynamics involving some downstream targets. The receptor production was modelled as the gene \emph{ptc} expression with two different kinetics. The first one is a basal constant production and the second one is an input from CiA restricted by CiR, both non-linearly represented. These two terms were repressed by a non-linear En contribution. To attempt this simplification it is proposed a constant number of receptors \cite{Umulis:2009ct} in each cell with a certain binding ($\gamma_{Ptc-Hh}$) and unbinding ($\gamma_{Ptc-Hh}^-$) rates. This way eq. (\ref{eq:ptc0}) for \emph{ptc} can be eliminated, the translational rate $\theta_{ptc}=0$ in eq. (\ref{eq:pPtc0}) and eqs. (\ref{eq:Hh0}, \ref{eq:pPtc0} and \ref{eq:PtcHh0}) can be written as follows:

\begin{equation}\label{eq:Hh00}
\frac{\partial Hh}{\partial \tau }=D\frac{\partial^2 Hh}{\partial x^2 }+\delta(x)\alpha_{hh}-\gamma_{Ptc-Hh}Ptc\cdot Hh-\beta_{Hh}Hh+\gamma_{Ptc-Hh}^-PtcHh
\end{equation}

\begin{equation}\label{eq:pPtc00}
\frac{\partial Ptc}{\partial \tau }=-\gamma_{Ptc-Hh}Ptc\cdot Hh-\beta_{Ptc}Ptc+\gamma_{Ptc-Hh}^-PtcHh
\end{equation}

\begin{equation}\label{eq:PtcHh00}
\frac{\partial PtcHh}{\partial \tau }=\gamma_{Ptc-Hh}Ptc\cdot Hh-\beta_{PtcHh}PtcHh-\gamma_{Ptc-Hh}^-PtcHh
\end{equation}

The result of this simplification is a replication of the ocellar pattern, as given in the previous model, for every species involved except for Ptc and PtcHh due to the absence of En interaction. This modification reveals the need to preserve the interaction between En and the receptor dynamics as, in order to maintain the full biological sense, PtcHh needs to show the correct patterning. Note that this will be regulated in a subsequent modification to the model by re-inserting En repression and CiA feedback.

The next downstream step in the network is the activation of \emph{ci} dependent proteins. In the former model, the gene $ci$ is modeled to be expressed dependent on a non-linear En repression, eq. (\ref{eq:ci0}). Then, CiA is produced with a linear dependence on $ci$ expression, eq. (\ref{eq:CiA0}), which is turned into CiR form inversely dependent on the ratio of bound and unbound Ptc to Hh, eqs. (\ref{eq:CiA0}-\ref{eq:CiR0}). In a coarse-grain view of this process, the activation of CiA correlates with Hh signalling inside the cell, and so, with the amount of Hh bound to its receptor.  To perform this simplification it is possible to go through different small steps which all maintain the desired pattern. Firstly, eq. (\ref{eq:ci0}) can turn from a constant production repressed by En to a Hh-like profile repressed by En when inserting a $PtcHh/Ptc$ contribution:

\begin{equation}\label{eq:ci1}
\frac{\partial ci}{\partial \tau }=\kappa_0\,\beta_{ci}\left(\alpha_{ci}\,\phi(\frac{PtcHh}{Ptc}\psi(En,k_{En-ci},n_{En-ci}),k_{ci},n_{ci})-ci\right)
\end{equation}

This means that the \emph{ci} expression is dependent on both the concentration of receptors bound to ligands and free receptors, which is a plausible result to be experimentally tested. Note that this interaction in the real system could be done either direct or indirect. In the latter case, an unknown species downstream the receptor dynamics could sense both, the concentration of free and bound receptors and, accordingly use this information to activate \emph{ci} expression.

Now, as $ci$'s output is already in a Hh-like profile in eq. (\ref{eq:ci1}), one can assume that CiA is self-stable and thus neglect the transformation from CiA to CiR in eqs. (\ref{eq:CiA0}-\ref{eq:CiR0}), and so, eq. (\ref{eq:CiR0}) for CiR can be eliminated. CiA dynamics is shown in eq. (\ref{eq:CiA1}).

\begin{equation}\label{eq:CiA1}
\frac{\partial CiA}{\partial \tau }=\kappa_0\,\beta_{CiA}\left(\theta_{ci}\,ci-CiA\right)
\end{equation}

At this point, the system stands for another change in a coarse-grain view. One can avoid the description of the receptor-ligand binding process if the downstream targets maintain their profile, as done in \cite{Zhou:2012ia}. As it has already been commented, $ci$ spatial form correlates with PtcHh and Hh profiles. Therefore, the next proposed simplification assumes that the bound Hh can behave as an internalized molecule dependent on Hh concentration instead of following the binding process. This, in the real system, can stand for the description of a molecule situated downstream of PtcHh and upstream of $ci$ in the Hh-signalling pathway. This other molecule senses the effect of Hh and can be altered by other elements as En and CiA. To simplify notation, PtcHh is used as this internalized molecule, now sensitive to En repression. This analysis resumes the need of including En interaction to the PtcHh species commented above. With this assumption, the first part of the pathway is modeled as follows:

\begin{equation}\label{eq:Hh01}
\frac{\partial Hh}{\partial \tau }=D\frac{\partial^2 Hh}{\partial x^2 }+\delta(x)\alpha_{hh}-\alpha_{PtcHh}PtcHh-\beta_{Hh}Hh
\end{equation}

\begin{equation}\label{eq:PtcHh01}
\frac{\partial PtcHh}{\partial \tau }=\alpha_{Hh-PH}\phi(Hh,k_{Hh},n_{Hh})\phi(CiA\,\psi(En,k_{En-PH},n_{En-PH}),k_{CiA-PH},n_{CiA-PH})-\beta_{PtcHh}PtcHh
\end{equation}

Note that an equation for unbound Ptc can be eliminated from the system as the binding process is no longer described. Most of the constant basal productions were set to zero in the original model. It has also been tested that the system keeps the same behavior when neglecting them so they are removed from the equations to simplify the equations structure at this point of the network pruning.

Now a straight forward simplification is performed to the system. The original model showed a detailed dynamics involving both genes expression and protein dynamics. One can notice that in the steady state, the genes expression and the protein concentration differ in a proportionality constant. According to this, one can represent the system using just one of the element types adjusting this proportionality constant in each case. The system is then shown in eqs. (\ref{eq:Hh01}-\ref{eq:Hth01}).

\begin{equation}\label{eq:CiA02}
\frac{\partial CiA}{\partial \tau }=\kappa_0\,\beta_{CiA}\left(\theta_{ci}\,\alpha_{ci}\,\phi(PtcHh\,\psi(En,k_{Enci},n_{Enci}),k_{ci},n_{ci})-CiA\right)
\end{equation}

\begin{equation}\label{eq:En01}
\frac{\partial En}{\partial \tau }=\theta_{en}\left(\phi(CiA\psi(CiR,k_{CiR},n_{CiR}),k_{CiA},n_{CiA})+\kappa_{En}\,\phi(En,k_{DlEn},n_{En})\right)-\beta_{En}\,En
\end{equation}

\begin{equation}\label{eq:Eya01}
\frac{\partial Eya}{\partial \tau }=\theta_{eya}\left(\alpha_{Toy}\,\phi(CiA\psi(Hth,k_{Hth-eya},n_{Hth}),k_{CiA-eya},n_{CiA})+\phi(Eya,k_{Eya},n_{Eya})\right)-\beta_{Eya}\,Eya
\end{equation}

\begin{equation}\label{eq:Hth01}
\frac{\partial Hth}{\partial \tau }=\theta_{hth}\left(\alpha_{Wg}\,\phi(\psi(Eya,k_{Eya-hth},n_{Eya}),k_{Wg},n_{Wg})\right)-\beta_{Hth}\,Hth
\end{equation}

The proposed GRN to fulfill the ocellar complex pattern is shown in (Fig. \ref{fig:1}d). 
A sum up description of this network is the following. The morphogen Hh diffuses extracellularly and forms the input profile of the system. Hh enhances an intracellular PtcHh concentration which causes a decrease in Hh concentration.
Then, PtcHh results in signaling for Cubitus Interruptus (CiA) in its active form. 
CiA activates the inhibitor Engrailed (En) that applies negative feedback loops at two different points on the Hh signaling pathway, PtcHh and CiA.
Finally CiA activates the formation of Eya which interacts with Homothorax (Hth) in a switch-like form, so Hth is present where Eya is not, and vice versa. Hth is activated by Wingless (Wg). Though Wg is a well known morphogen, in these models it is considered as a constant input. Wg is, then, not behaving as a graded morphogen but a constant input. This takes into account that its effect in the system would be qualitatively similar if its expression is flat in the whole imaginal disc \cite{Alexandre:2014fa}. After taking the simplifications where many of the interaction mechanisms are referred in a coarse-grain description, we will refer hereinafter to Eya as RD proteins, without particularly considering any of them but showing a general overview of the RD complex dynamics. Then, the spatial domains in the model where RD is present correspond to the ocellar cuticle. (See complete results in Supplementary Material).

\subsubsection{Changes in formulation paradigm maintain the correct patterning}
In order to perturb the structure of the model's equations, the system has been substantially pruned from that in \cite{AguilarHidalgo:2013bw}. Now one can question whether the patterning can be affected by the formulation used. In order to overcome this issue, the model has been re-written using a mostly linear formulation (eqs. (\ref{eq:deltax}) to (\ref{eq:Deltaz})). This way, Hill's function formulation or any other kind of relaxation kinetics used to simulate more realistic biological behaviors are avoided except in necessary cases (see Supplementary Material).
%Then, in order to keep positive all the concentrations, Heaviside functions $\Theta$ are used instead. 

%The ocellar complex physical model is formulated by the following 6 differential equations of the reaction-diffusion type and 2 functions.
\begin{equation} \label{eq:deltax}
\delta(x) = \left\{
	       \begin{array}{ll}
		 1      & \textmd{if\ } x \in \textmd{\emph{hh}-expressing cells} \\
		 0      & \textmd{if\ } x \notin \textmd{\emph{hh}-expressing cells} \\
	       \end{array}
	     \right.
\end{equation}

\begin{equation}\label{eq:Hh}
\frac{\partial Hh}{\partial \tau }=D\frac{\partial^2 Hh}{\partial x^2 }+\delta(x)\alpha_{hh}-\alpha_{PtcHh-Hh}PtcHh-\beta_{Hh}Hh
\end{equation}

\begin{equation}\label{eq:PtcHh}
\frac{\partial PtcHh}{\partial \tau }=\Theta\left(\alpha_{Hh-PtcHh}Hh+\alpha_{CiA-PtcHh}CiA-\alpha_{En-PtcHh}En\right)-\beta_{PtcHh}PtcHh
\end{equation}

\begin{equation}\label{eq:CiA}
\frac{\partial CiA}{\partial \tau }=\frac{\frac{\gamma_{PtcHh-CiA}}{1+\alpha_{En-CiA}En}\alpha_{PtcHh-CiA}PtcHh^{n_{CiA}}}{k_{m_{CiA}}^{n_{CiA}}+\alpha_{PtcHh-CiA}PtcHh^{n_{CiA}}}-\beta_{CiA}CiA
\end{equation}

\begin{equation}\label{eq:En}
\frac{\partial En}{\partial \tau }=\frac{\gamma_{CiA-En}CiA^{n_{En}}}{k_{m_{En}}^{n_{En}}+CiA^{n_{En}}}+\alpha_{En-En}EnDl(\zeta )-\beta_{En}En
\end{equation}

\begin{equation}\label{eq:RD}
\frac{\partial RD}{\partial \tau }=\Theta\left(\alpha_{CiA-RD}CiA-\alpha_{Hth-RD}Hth+\alpha_{RD-RD}RD\right)-\beta_{RD}RD
\end{equation}

\begin{equation}\label{eq:Hth}
\frac{\partial Hth}{\partial \tau }=\Theta\left(\alpha_{Wg-Hth}-\alpha_{RD-Hth}RD\right)-\beta_{Hth}Hth
\end{equation}

\begin{equation} \label{eq:Deltaz}
Dl(\zeta_{En},\zeta_{\tau},\tau_{\zeta_{En}}) = \left\{
	       \begin{array}{ll}
		 Dl     & \textmd{if } En \geq \zeta_{En},\, \forall \tau \in \left[ \tau_{\zeta_{En}},\,\tau_{\zeta_{En}}+\zeta_{\tau}\right] \\
		 0      & \textmd{otherwise} \\
	       \end{array}
	     \right.
\end{equation}
\\
The coefficients $\alpha_{X-Y}$ drive an action (activation or repression, depending on sign, $+$ or $-$ respectively) from source element X to target element Y; $\beta_X$ is the degradation coefficient of the element X; $\gamma_{X-Y}$ is the maximum Hill term value when X regulates Y.

\begin{figure}
\begin{center}
\includegraphics[width=\textwidth]{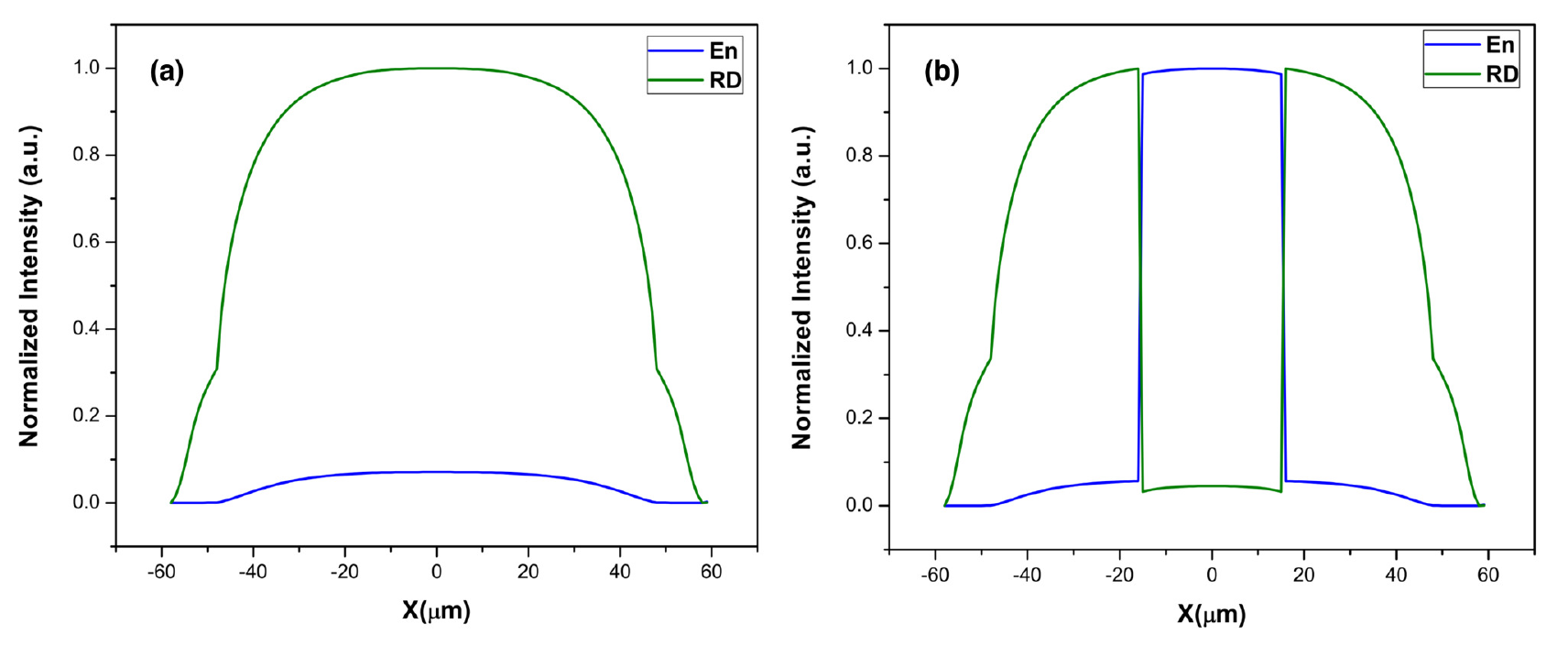}
\end{center}
\caption{(a) Representation of RD and En normalized profiles in the case that En is not self-regulating. In this case, the RD is present in the whole ocellar complex appearing as one unsplit domain. The parameter values used to obtain this result are shown in Table 2 of Supplementary Material, modifying $Dl$ parameter value to $Dl=0$. (b) Representation of RD and En normalized profiles in case En is self-regulated just in the cells where its concentration is over a determined threshold ($Dl=1$).  Now RD defines the ocellar domains while En characterizes the IOC.}
\label{fig:2}
\end{figure}

The proposed linear formulation is not conservative so some parameter sets could lead to negative concentrations. To avoid this issue Heaviside functions $\Theta$ are introduced. In this new system, equations (\ref{eq:Hh}-\ref{eq:Hth}) are the reaction-diffusion type differential equations of the model. The description of the system is the same as in the last section. Some non-linearities are applied in order to maintain the correct pattern:
%Equation (\ref{eq:Hh}) is the Hh equation. Hh is produced just in the central cells (eq. (\ref{eq:deltax})) and its concentration is reduced when increasing PtcHh's and by its own degradation rate.  Equation (\ref{eq:PtcHh}) shows the behavior of PtcHh element. Here we are considering PtcHh not as a complex formation but as a different species which concentration increases with a cost in Hh concentration, simplifying the complex formation description. Also, PtcHh activation is favored by CiA and repressed by En. CiA dynamics is formulated in equation (\ref{eq:CiA}). Here, CiA is activated by PtcHh and repressed by En. 
Firstly, En repression of CiA is shown as a reduction of the maximum level that CiA can reach ($\gamma_{PtcHh-CiA}$). A stability study shows that this kind of non-linear repression makes this system have a stable point, while linear repression leaves the system without any stable solution (see Supplementary Material). 
Secondly, in order to achieve the desired pattern, En needs to be maintained in a Hh signaling-independent manner. As CiA is the only En activator, En should self-maintain. This new behavior should not work in every cell, as the result would be a complete En domain with no response to Hh influence (Fig. \ref{fig:2}a); so the En self-maintenance should work only under certain circumstances, and these are Hh signaling-dependent. In order to treat this feature in a simplistic way En is self-maintained if its concentration exceeds a certain threshold, $\zeta_{En}$, for a time interval, $\zeta_{\tau}$, starting from the time point when En reaches the concentration threshold, $\tau_{\zeta_{En}}$, (eqs. \ref{eq:En} and \ref{eq:Deltaz}). This requirement refers to the need of integrating a protein quantity over time to trigger some targets \cite{Dessaud:2010km}. 
%It has been determined that Delta/Notch signaling pathway (Dl/Notch) is required for En self-maintenance \cite{AguilarHidalgo:2013bw}. 

These premises form an En domain in the IOC allowing the production of the RD proteins in the OC region (Fig. \ref{fig:2}b). If En transcription is eliminated ($\gamma_{CiA-En} = 0$) or substantially reduced so that it cannot reach enough concentration to self-regulate, then RD appears in a single unsplit domain that follows the Hh profile. There is experimental evidence showing that at an early developmental stage (mid L3) RD is appears as a single unsplit domain, similar to the model's output. However, this pattern evolves so that by the end of L3, RD expression and Hh signaling are detected in two domains. These domains are separated by a region in which RD and Ptc expressions are very low or absent \cite{Dessaud:2010km,AguilarHidalgo:2013bw}. This pattern suggested the existence of a repressor capable of shutting down the \emph{hh} pathway. 
%This confirms the need for incorporating En repressor into the system. 
The same pattern is observed if Dl signaling were shut off ($Dl = 0$), the mathematical effect would be identical, En self-regulation term will not cause any effect and RD will appear in only one domain over the whole tissue. 
%RD is activated by CiA and also self-regulates (eq. 6). RD has a repressor, Hth, which is also repressed by RD itself in a mutual double repression form (eqs. \ref{eq:RD},\ref{eq:Hth}). Hth is activated by Wingless (Wg) that in this model is treated as a constant production term $\alpha_{Wg-Hth}$. 

%; $\zeta_{En}$ is the En value threshold from which En self-regulates, the same way $\zeta_{\tau}$ is the time threshold from which En is able to self-regulate only if En concentration exceeds $\zeta_{En}$ (eq. \ref{eq:Deltaz}). 

\begin{figure}
\begin{center}
\includegraphics[width=0.7\textwidth]{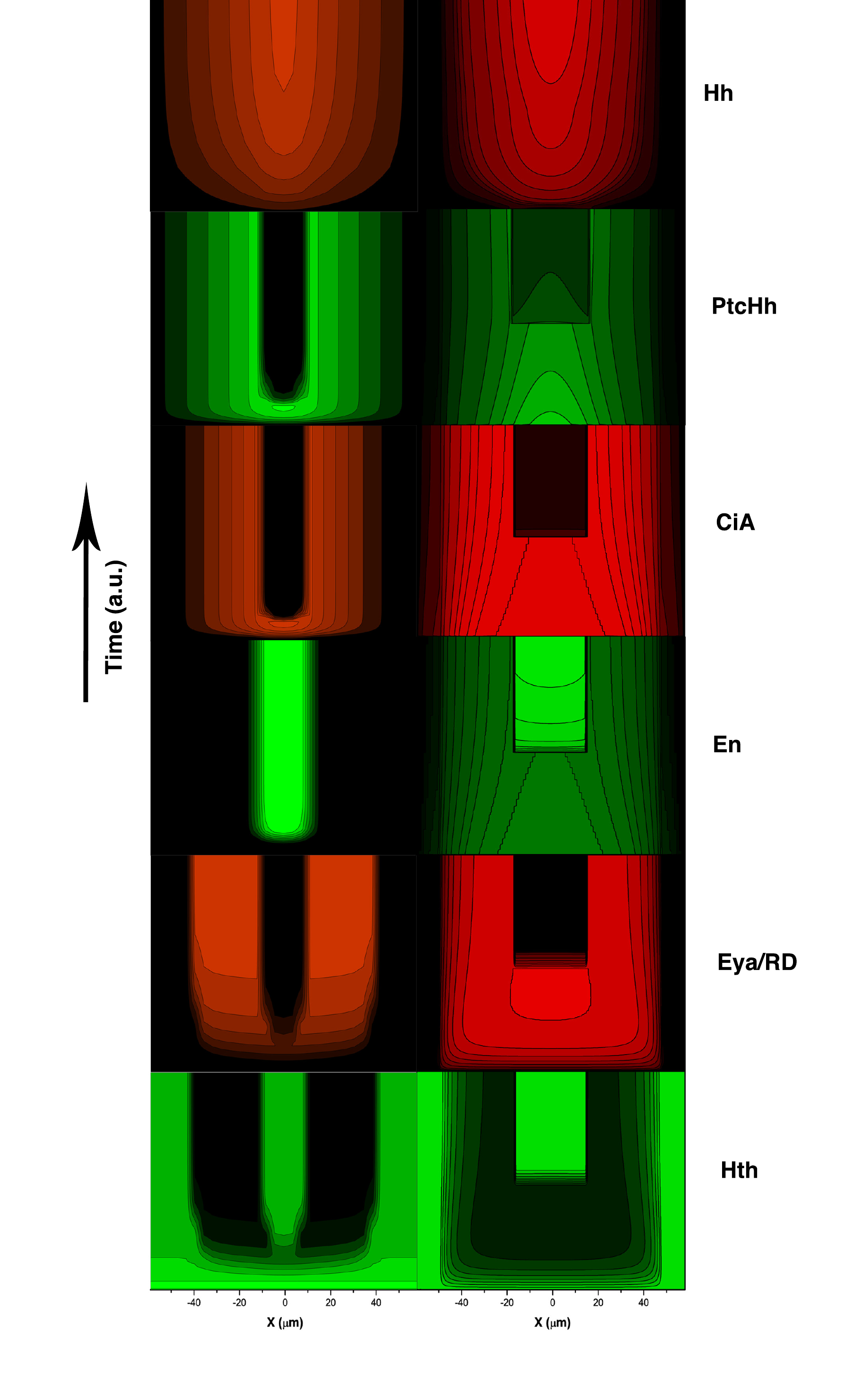}
\end{center}
\caption{Spatio-temporal representation of the signal intensity of the non-linear (left - eqs. (\ref{eq:Hh01}-\ref{eq:Hth01})) and linear (right - eqs. (\ref{eq:Hh}) to (\ref{eq:Hth}))) simplified models. From up to down: Hh, PtcHh, CiA, En, Eya/RD, Hth. The brighter the color, the higher the concentration value.}
\label{fig:3}
\end{figure}

The pruned GRNs both linearly and non-linearly formulated are able to reach the correct pattern formed by two OC domains separated by the IOC (Fig. \ref{fig:3}). In this interocellar region Eya/RD values are null or very low while En concentration is high; likewise, in the ocellar domains En concentration is null or very low and RD concentration is high. The regulation of the different domains depends on the Hh gradient concentration value on each spatial point. En concentration is dependent on CiA concentration which also depends on Hh concentration as output of its signaling pathway. Therefore, the spatial point where En is starting self-regulating depends on Hh concentration. This behavior responds to the classical \emph{French flag model}.
It is also noteworthy that Hth is only present where RD is not, due to the double negative feedback control between these two elements. So Hth shows high concentration in the interocellar and periocellar regions.
PtcHh and CiA behaviors are also shown as expected. They reflect the Hh gradient as being part of the Hh-signaling pathway and feel the negative effect of En repression in its domain (IOC). One must note that the IOC domain is set in a more abrupt way in the linearly formulated model than in the non-linearly done one. This refers to the use of a step function for the En self-regulation (linear model) instead of a Hill's equation (non-linear model). As in \cite{AguilarHidalgo:2013bw}, the simplified model is robust against noise perturbations. This model also shows a size control of the OC when perturbing the activation of Hth by Wg (see the details of these results in Supplementary Material).

\subsection{A 3-nodes network motif retrieves the patterning: The core}

\begin{figure}
\begin{center}
\includegraphics[width=\textwidth]{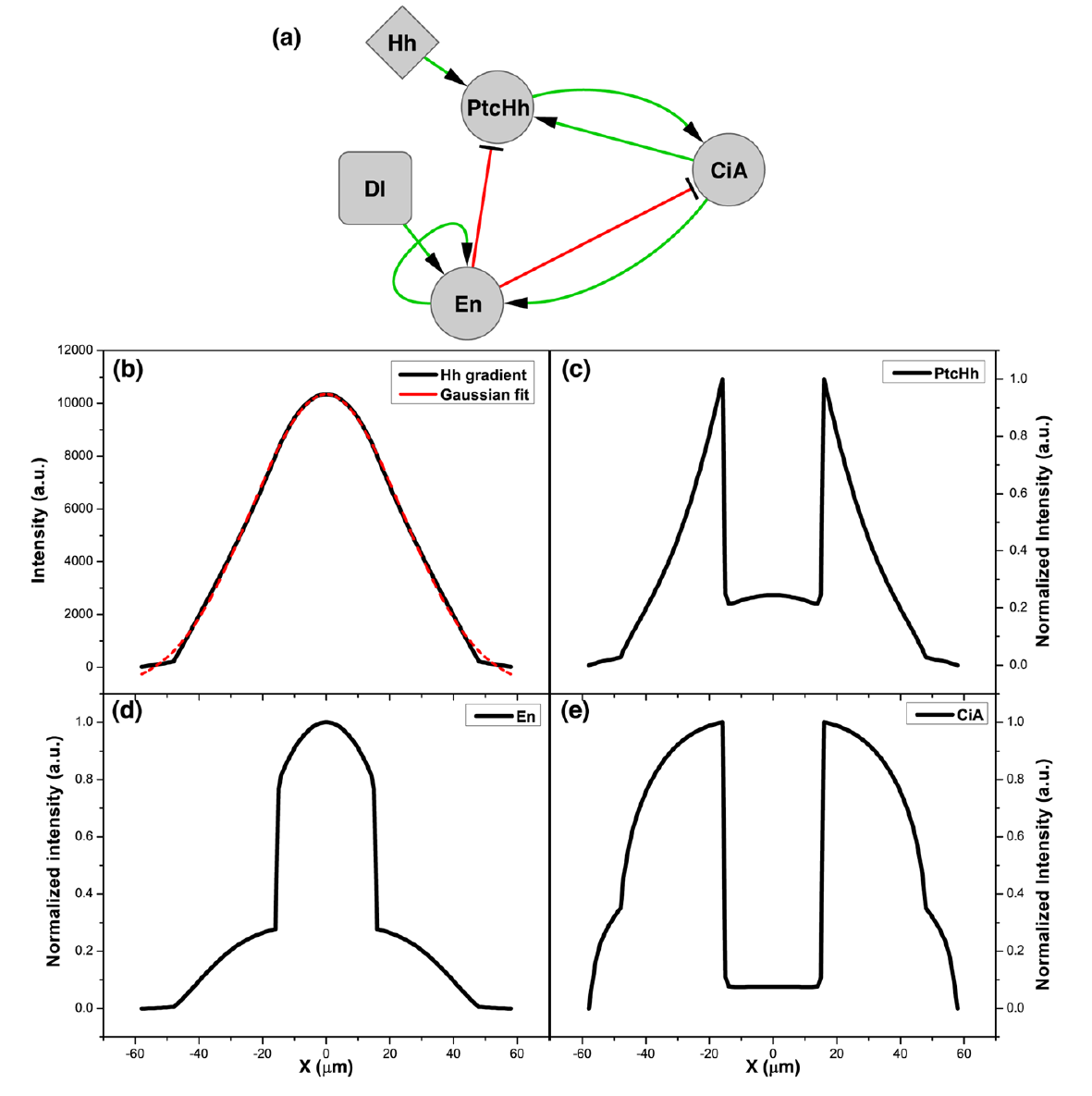}
\end{center}
\caption{(a) Network diagram of the 3-nodes model. The network dynamics is the following: Hh is the input to the network in the form of a static Gaussian profile that activates PtcHh. PtcHh activates CiA while CiA activates both PtcHh and En. En represses both PtcHh and CiA. En is self-maintained in the case Dl function is activated. Dl activates En self-regulation if En surpasses a certain concentration. (b) Hh gradient steady state in the model from the pruned GRN (in black line) and its Gaussian fit (in dashed red line). (c-e) Solutions of the 3-nodes model of 3 equations for (c) PtcHh, (d) En and (e) CiA.}
\label{fig:7}
\end{figure}

The study made to this point shows that the correct patterning seems to be extremely correlated to the use of a negative feedback loop with a dependence on the morphogen concentration to trigger this inhibition. To finally prove this, a single 3-node network motif has been extracted maintaining the negative feedback loop. The description of this motif is the following: PtcHh activates CiA, CiA activates both En and PtcHh, and En represses the previous two network nodes in a double negative feedback loop (Fig. \ref{fig:7}a). The input to this network motif is, in an adiabatic approximation, a stationary Hh gradient-like profile to simplify the calculations, avoiding this way the PDE equation for Hh. Both the analytical solution for equation (\ref{eq:Hh}) in its stationary state or its Gaussian fit (see Supplementary Material.), provide the same input to the system as in the dynamical case in a steady state situation. Fig. \ref{fig:7}b shows the comparison between the numerical stationary solution of the Hh morphogen gradient and the Gaussian distribution fit (see details in Supplementary Material). Then, the same system's behavior is maintained. By doing this, the system becomes a system of ODEs of the reaction type.

%Up to this point, the 3-nodes network can be fed, in an adiabatic approximation, by a static Hh profile instead of the dynamical Hh diffusion equation (\ref{eq:Hh}). Both the analytical solution for equation \ref{eq:Hh} in its stationary state or its Gaussian fit (see supp. mat.), provide the same input to the system as in the dynamical case, so then the same system's behavior is maintained. By doing this, the system becomes a system of ODEs of the reaction type.
The resulting system is formed then by three equations.

\begin{equation}\label{eq:PtcHh2}
\frac{\partial PtcHh}{\partial \tau }=\Theta\left(\alpha_{Hh-PtcHh}\,Hh(x)+\alpha_{CiA-PtcHh}\,CiA-\alpha_{En-PtcHh}\,En\right)-\beta_{PtcHh}\,PtcHh
\end{equation}

\begin{equation}\label{eq:CiA2}
\frac{\partial CiA}{\partial \tau }=\frac{\frac{\gamma_{PtcHh-CiA}}{1+\alpha_{En-CiA}\,En}\alpha_{PtcHh-CiA}\,PtcHh^{n_{CiA}}}{k_{m_{CiA}}^{n_{CiA}}+\alpha_{PtcHh-CiA}\,PtcHh^{n_{CiA}}}-\beta_{CiA}\,CiA
\end{equation}

\begin{equation}\label{eq:En2}
\frac{\partial En}{\partial \tau }=\frac{\gamma_{CiA-En}\,CiA^{n_{En}}}{k_{m_{En}}^{n_{En}}+CiA^{n_{En}}}+\alpha_{En-En}\,En\,Dl(\zeta_{En},\zeta_{\tau},\tau_{\zeta_{En}})-\beta_{En}\,En
\end{equation}

These equations (\ref{eq:PtcHh2}-\ref{eq:En2}) have the same formal expression as in eqs. (\ref{eq:PtcHh}) (PtcHh), (\ref{eq:CiA}) (CiA) and (\ref{eq:En}) (En) but considering a constant Hh profile in PtcHh equation (Fig. \ref{fig:7}b).

This 3-nodes ocellar model is still capable of showing qualitatively identical expressions for PtcHh and En (Figs. \ref{fig:7}c and \ref{fig:7}d respectively) and the correct ocelli pattern in CiA element (Fig. \ref{fig:7}e) without changing the parametric values from the former model. It is noteworthy to emphasize that both a dynamic morphogen profile and its stationary solution retrieves the correct pattern. Thus, one can provide two important results. The first one implies the use of the dynamical description. Here, the behavior of the system is positively described at different developmental stages when generating the Hh profile from null concentration values (as seen in previous sections). The second result comes from the solution of the patterning when using an adiabatic approximation to the morphogen dynamics. Here, one can obtain the correct pattern when using a steady graded profile as the system's input. In other words, the ocellar patterning can be fulfilled also in case the morphogen kinetics is much faster than the intracellular response.

\subsection{Reduction of the system to an activator-inhibitor dynamics}\label{sec:act-rep}
The 3-nodes network model raises the question of whether an ultimate simplification on the system will lead to a correct patterning. In this case, though biologically unrealistic, it is considered that Hh is directly feeding CiA production with no intermediate; CiA is then activating En and this last represses CiA formation. The correct pattern is still maintained. It is clear that this \emph{French flag} type model forms the ocellar pattern due to this activator-inhibitor interaction. The patterning is correctly achieved if this specific set is maintained in the core of the network.

Following the 3-nodes model formulation it is possible to write this system in terms of ODEs as follows:

\begin{equation}\label{eq:CiA_2eq}
\frac{\partial CiA}{\partial \tau }=\frac{\gamma_{Hh-CiA}\,Hh(x)}{1+\alpha_{En-CiA}\,En}-\beta_{CiA}\,CiA
\end{equation}

\begin{equation}\label{eq:En_2eq}
\frac{\partial En}{\partial \tau }=\frac{\gamma_{CiA-En}\,CiA^{n_{En}}}{k_{m_{En}}^{n_{En}}+CiA^{n_{En}}}+\alpha_{En-En}\,En\,Dl(\zeta_{En},\zeta_{\tau},\tau_{\zeta_{En}})-\beta_{En}\,En
\end{equation}

This simple system allows an analytical treatment to find the steady state solutions for En and CiA. In order to ease the analysis, the steady state solutions for En and CiA in the steady state are written in a dependent manner:

\begin{equation}\label{eq:CiA_2eqss}
CiA=\frac{1}{\beta_{CiA}}\frac{\gamma_{Hh-CiA}\,Hh(x)}{1+\alpha_{En-CiA}\,En}
\end{equation}

\begin{equation}\label{eq:En_2eqss}
En=\frac{1}{\beta_{En}-\alpha_{En-En}\,Dl(\zeta_{En},\zeta_{\tau},\tau_{\zeta_{En}})}\frac{\gamma_{CiA-En}\,CiA^{n_{En}}}{k_{m_{En}}^{n_{En}}+CiA^{n_{En}}}
\end{equation}

A simple observation of eq. (\ref{eq:En_2eqss}) shows that in the case of En self-regulation, this is $Dl(\zeta_{En},\zeta_{\tau},\tau_{\zeta_{En}})=1$, if the parameters $\alpha_{En-En}$ and $\beta_{En}$ are equal or very similar, En concentration goes to infinity or to a very high value; and this leads to a null or very low CiA value. So the Hh signaling pathway is closed and no ocellar cuticle is developed. In the case of $Dl(\zeta_{En},\zeta_{\tau},\tau_{\zeta_{En}})=0$, depending on the parameter values, positive or imaginary values are obtain for both En and CiA. Though imaginary solutions can be found if $n_{En}>1$, only real solutions are considered, and these are positive. So in this case,  CiA expression is maintained and OC is formed.

\subsection{Boolean networks analysis confirms robustness in the patterning circuit}\label{sec:boolean}

It has been demonstrated that several formulations of the same GRN with different levels of complexity are able to determine the same patterning. This pattern seems indeed to be robust to the choice of model if the activator-repressor module is maintained in the network. In this section, in order to test the robustness of this pattern independently of the parameter values and the dynamics of a mean-field formulation, the GRN has been studied from the perspective of Boolean networks. For this, four models have been tested, the one in Fig. \ref{fig:1}c for the model in \cite{AguilarHidalgo:2013bw} (model $\#1$), the one in Fig. \ref{fig:1}d (model $\#2$), the described in Fig. \ref{fig:7}a (model $\#3$) and the activator-repressor motif network detailed in \cref{sec:act-rep} (model $\#4$).

To perform the Boolean models, the state of each node is 1 or 0, depending on the presence or not of the corresponding genetic element. The states of the nodes can change in time, and the next state of node $i$ is determined by a Boolean (logical) function $\Phi_i$ of its state and the states of those nodes that have incoming edges on it. In general, a Boolean or logical function is written as a statement acting on the inputs using the logical operators \emph{and}, \emph{or} and \emph{not} and its output is $1(0)$ if the statement is true (false). For the construction of the $\Phi_i$ functions it is assumed that the inhibitors are dominant in the networks, so they will be treated as AND logical operators \cite{Albert:2003ic}.
This type of network modeling has already been widely used to study genetic networks \cite{Kauffman:1969up,Thomas:1973ve,Thomas:1990tg,Kauffman:1993uj,Albert:2003ic,Gonzalez:2006bf,Buceta:2007bg}. Table \ref{tab:boolean} gives the Boolean functions $\Phi_i$ for every genetic element of the four considered models.

\begin{figure}
\begin{center}
\includegraphics[width=\textwidth]{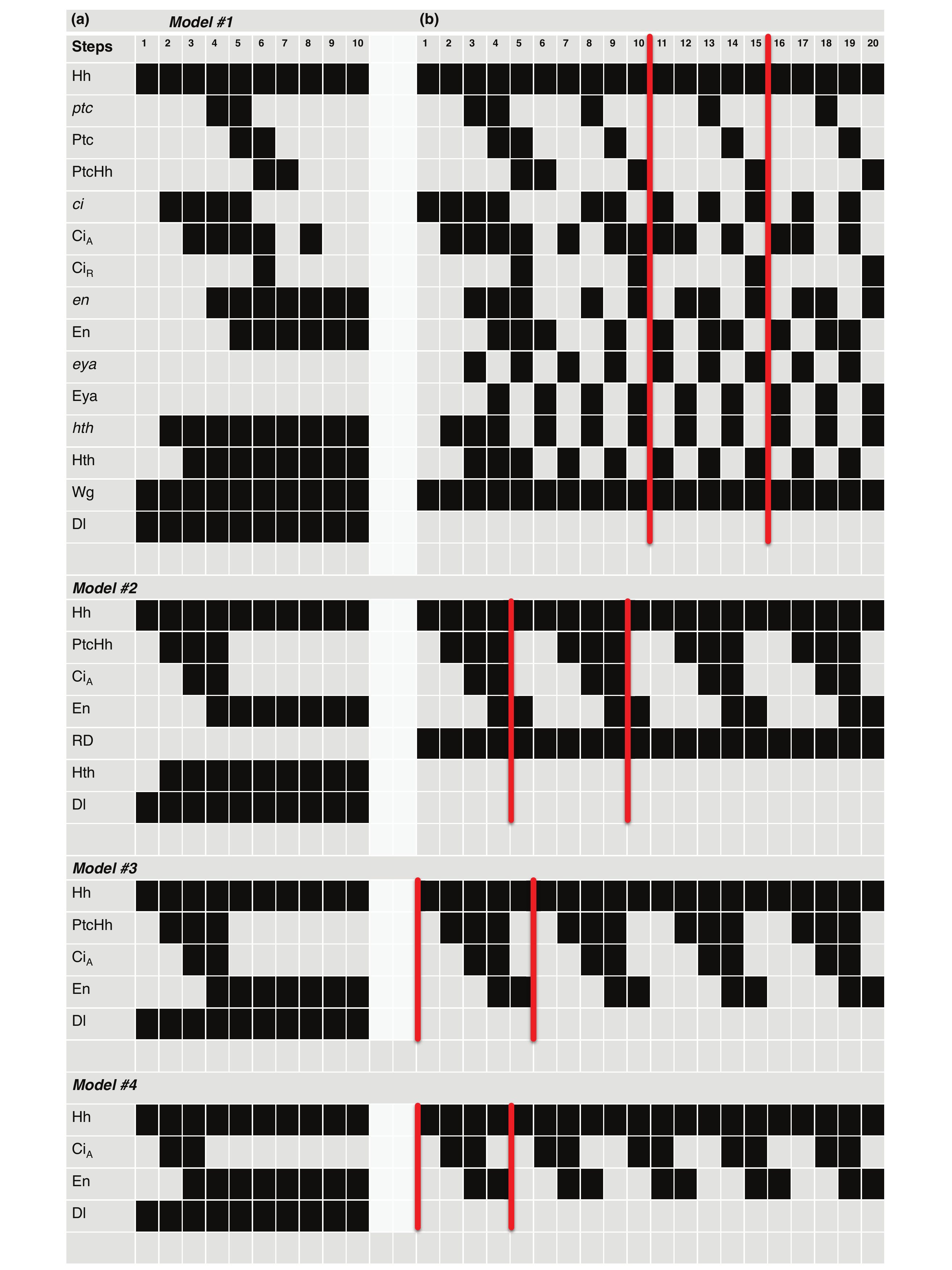}
\end{center}
\caption{Time series showing the behavior of the different network models from their Boolean perspective (black = 1, grey = 0). (a) Situation in which the Dl/Notch signaling pathway trigger the En self-regulation. The four models show a steady-state solution where the Hh-signaling pathway is blocked by En defining the IOC. (b) Scenario where En is not self-regulated and the OC region is defined by CiA and Eya in an oscillatory regime. Only model \#2 shows a steady situation for RD as Hth persistently represses it. Red lines show the oscillation period.}
\label{fig:8}
\end{figure}

Models $\#1$, $\#2$, $\#3$ and $\#4$ show the same behavior. In case that the network is describing the behavior of a cell in the IOC region (Fig. \ref{fig:8}a), Dl = 1 so En self-regulates, the Hh-signaling pathway is blocked and the RD genes are not expressed. This situation is given as a steady state regime in the four analyzed models. It is found the steady state takes longer to be reached as both the number of elements involved in the system and their relationships increase. In model $\#1$, the active nodes in the steady state are Hh, $en$, En, $hth$, Wg, Hth and Dl. Model $\#2$ shows the same behavior so the only active nodes are Hh, En, Hth and Dl. In models $\#3$ and $\#4$ the active nodes are then Hh, En and Dl. The IOC is well defined in the four models.
In the case of the OC region (Fig. \ref{fig:8}b), Dl is set to 0 and En does not self-regulate. Interestingly, this state in the four models also show the same behavior though reveals an oscillatory regime. One can observe that CiA and RD (depending on the studied model) are active in time, so the ocellar pattern is fulfilled though in a non-steady situation. Models $\#1$ and $\#2$ show a transient, that is longer in the first one. Models  $\#1$, $\#2$ and $\#3$ show the same period length of 5 time steps while model $\#4$ present a shorter period of 4 time steps, this is due to the reduction of the number of nodes between the input, Hh, and the repressor feedback. 

Models $\#1$ and $\#2$ show a dependence on the initial conditions to express a different behavior in nodes related to the outcome of the network. These are $eya$, Eya, $hth$ and Hth in model $\#1$, and RD and Hth in model $\#2$. This dependence is due to the mutual double repression between the Eya/RD and Hth species. The switch-like mechanism will activate only one of the two family species or both in an oscillatory fashion depending on the initial conditions given. In Fig. \ref{fig:8} is shown the behavior where the only active node is Hh as initial condition for models $\#3$ and $\#4$. For model $\#1$, Wg is also considered active from the initial step. In the OC description of this model, $ci$ has also been set active as initial condition in model \#1 to allow the Eya and Hth species rule in an oscillatory manner. If $ci$ is set inactive as initial condition, the Eya species are set to 0 as Wg is always 1 and it activates Hth before Eya can be turned on. Therefore, the switch is kept this way in time. In model \#2, in order to maintain the switch always activating the RD species it is necessary to initialize RD as 1. If this is not considered, Hth will finally block RD in a persistent way as the oscillation period of CiA is larger than the number of links between RD-Hth in this model, so even initializing $\text{CIA}=1$ Hth will block RD if RD is 0 in its initial state. From this one can extract that, in the real network, either the activation of RD species is done earlier than Hth's in the developmental program, or the sub-network involving RD and Hth species is more complex than a simple mutual interaction. The full analysis of the oscillatory regime and the dynamical properties of the network species involved will be released by the authors in another work.

\section{Materials and Methods}

The total one-dimension ocellar complex is built on a linear lattice of 117 nodes. It has been measured for this work that the IOC and each ocelli is approximate $30 \mu m$ long. The ocellar complex is approximately 30 cells long and the cell diameter is $\sim 4 \mu m$. A good approximation in the model is to set each lattice node as a $1 \mu m$ of tissue length. So, it can be assumed that each element can reach 33 nodes long, thus leaving 9 nodes free on each side for fitting purposes. This way, a linear space of 117 nodes is enough to consider the development of a satisfactory ocellar complex pattern. The spatial dimension on the graphical results are, then, shown in $\mu m$ units. 

A simplified one-node model (for the six equations (\ref{eq:Hh}-\ref{eq:Hth}) case) was implemented using Vensim software, a visual tool for solving ODEs that allows parameter values modification in run-time (Vensim PLE version 5.11, Ventana Systems, \emph{http://www.vensim.com/software.html}). This program contains a fourth order Runge Kutta method (RK4) to solve ODE systems. 

The 117-nodes models were also implemented using MATLAB (MathWorks) and solved with the integrator ode45. 

For all the numerical experiments, the differential equations system has been discretized using a finite differences method.
 
The values for all the parameters used in the simulations and the initial conditions for each variable can be found respectively in tables 1, 2 for the non-linear models and 3 and 4 for the linear models in the Supplementary Material. Some parameter values are cited and others have been mathematically derived. The rest of the parametric set has been manually adjusted to obtain a similar ocellar pattern to the experimental one.

The Gaussian fitting for Hh steady state profile was performed by Microcal Origin software.
The Boolean networks analysis was performed using application BooleSim \cite{Bock:2014hs}.

\section{Discussion}

How to establish a pattern in biological tissues is one of the most hot topics in recent developmental biology, though its interest has been shown from several decades from both experimental \cite{Wolpert:1994ei,Lum:2004gv,Dessaud:2010km,Kicheva:2007bha,Nahmad:2009cp,Wartlick:2011by,Restrepo:2014fw} and theoretical perspectives \cite{Collier:1996be,Meinhardt:1996vd,Barkai:1997cd,BenZvi:2010es,Nahmad:2010kc,Burda:2011hm,Nahmad:2011cb,Kicheva:2012ip,Li:2012vq}. Several mechanisms for pattern formation have been proposed and this research is still ongoing with many unanswered questions.
In the present work we are focusing on the robustness of the ocellar complex pattern in the \emph{Drosophila melanogaster} fly, in terms of perturbations in the system on the description level of the GRN structure representation. We started from a high description level PDE model using a non-linear formulation \cite{AguilarHidalgo:2013bw}.
% This study led us to find out of the GRN, the equations formulation and representation paradigm of the GRN. 
We have systematically modified this model by considering simpler assumptions, which keep the patterning, while maintaining the basic biology from a coarse-grain point of view. This study reveals that this pattern follows one of the most known patterning mechanisms, the \emph{French flag model} \cite{Sciences:1968wg}. This mechanism is driven by a morphogen gradient (Hh) where the cellular fate is determined by the concentration of this diffusive molecule, triggering different parts of the GRN.
The hh-signaling pathway is blocked by the expression of its inhibitor, En, which is also Hh-dependent. This inhibitor becomes Hh-independent when its concentration exceeds a certain threshold, self-regulating and repressing the expression of the RD genes. This way, at high Hh concentration, the En domain is set and the IOC is formed. Otherwise, En is not able to inhibit the expression of the RD genes and the OC region is fulfilled.
This basic mechanism seems to be robust in this GRN even after being substantially pruned. We have found that the patterning itself is independent, for example, on the receptor dynamics and on how the downstream signaling molecule CiA forms a graded profile. In this latter, we found that \emph{ci} expression could be directly or indirectly dependent on the concentration of receptor-bound morphogen.

In order to test if the patterning is significantly affected by the type of formulation used to simulate the model, we have implemented this model using both a non-linear and a linear paradigm. This result showed that the patterning seems to be also robust against the choice of formulation paradigm. Comparing the results of both types of formulation one can see that the non-linear model allows a precise control on the activation and repression processes by using Hill functions. This permits, for example, to completely block the Hh-signalling pathway by regulating the Hill function parameters on each regulatory term. This control is less clear in the linear formulation where the parameters contribute to the rate of change of concentration by adding/subtracting concentration dependant terms. On the contrary, the parameters in the linear model tell about effective rates that, in some cases, could be experimentally measured and give feedback to the model. One can note that the repression transition is smoother in the non-linear than in the linear formulation due to the use of a step function in the self-regulation of En in the linear case, eq. (\ref{eq:Deltaz}). We also found that the correct patterning is fulfilled when considering both a dynamical and a steady morphogen profile. In the dynamical description, the system can positively describe different developmental stages during Hh profile transition to its steady state from a null concentration scenario. On the other hand, as an adiabatic approximation also retrieves the correct pattern, we can infer that the morphogen dynamics is indeed not necessary to fulfill the pattern. This implies that the morphogen kinetics could be faster than the intracellular response so we can consider the morphogen dynamics as a serial of steady-state profiles.

By analyzing the minimal expression of the GRN capable of maintaining this behavior, we have found that the patterning is dependent on the specific topology of a core regulatory network motif containing an activator-repressor regulatory mechanism. This activator-repressor paradigm is a well known mechanism in pattern formation since \citet{Turing:1952vn}. In our case of study, the activator-repressor mechanism is local for each cell, representing a triggering structure for a classical \emph{French flag model}, where the activator-repressor network motif behavior responds to the morphogen concentration input in each cell. Though mathematically speaking, this activator-repressor mechanism is enough to fulfill the correct pattern by itself, biologically speaking, Hh cannot directly reach the activation of CiA but needs at least one intermediate to describe an extra-intracellular communication. This way, we consider the minimal representation of the GRN, with biological significance, the (core) 3-nodes sub-network described in this work. Here PtcHh is defined as the intermediate between Hh and CiA. This is, then, a network motif with a double negative feedback loop. It is also known that this type of three-component systems are appropriate to account for a wide class of biological phenomena \cite{Meinhardt:2004tc}. Some examples are found in the enrichment of microRNA targets \cite{Cui:2006ct,Marr:2010go}. A closer example to the system studied here is the developmental patterning of neural subtype specification in mouse \cite{Balaskas:2012cf}. There, the morphogen responsible for this patterning is the mouse Hh homologue Sonic Hedgehog (Shh). It is likely to think that this specific type of 3-nodes activator-repressor network motif can be responsible for the development of similar patterns in other developmental systems.

We have also confirmed the patterning independence on the dynamical model implementation complexity by analyzing the four studied network models from a Boolean perspective, and thus, the robustness of the GRN behavior and their patterning mechanism as long as the core network topology is maintained. We found that the IOC region is established in a steady state fashion. Interestingly, the OC region is defined as an oscillatory regime in the four models. Specifically, the activator-repressor mechanism is responsible for the emergence of these oscillations.
The inclusion of En as a repressor of the hh signaling-pathway establishes a negative feedback within the network that might be capable of changing the system dynamics non-linearly causing oscillations in gene expression. An experimental research for validating these oscillations during ocellar development has not been explored yet, though it is a tantalizing one. Oscillatory regimes have been proposed to increase the efficiency of reactions \cite{Richter:1980tr}, and to be involved in cell patterning by a morphogen graded profile \cite{Koch:1994un}. Even this oscillatory behavior could act as a switch-like mechanism between two transient cell-states building spatial borders among oscillating and non-oscillating cells.
\vspace{1cm}

\hspace{-0.5cm}\textbf{Acknowledgements}\\

 \hspace{-0.5cm}The authors would like to thank Fernando Casares (CABD) for introducing them to this fascinating system and for many fruitful discussions. This work is partially financed by Junta de Andaluc\'ia (FQM-122) to A. C\'ordoba and by the Spanish Ministry for Science and Innovation (MICINN/MINECO) and Feder Funds through grants BFU2012-34324 to F. Casares (CABD, Seville, Spain) and Consolider ÔFrom Genes to Shape,Õ of which F. Casares was a participant researcher.

\begin{table}[hbt]
\caption{Boolean functions}
\begin{tabular}{@{}ll@{}}
\hline\noalign{\smallskip}
Node  & Boolean updating function $\Phi_i$  \\
\noalign{\smallskip}\hline\noalign{\smallskip}
\emph{Model $\#1$} &\\
$Hh_i$ & $Hh_i = 1$ for all $t$\\
$ptc_i$ & $ptc_i^{t+1} = CiA_i$ AND NOT $CiR_i$ AND NOT $En_i$\\
$Ptc_i$ & $Ptc_i^{t+1} = ptc_i$\\
$PtcHh_i$ & $PtcHh_i^{t+1} = Hh_i$ AND $Ptc_i$\\
$ci_i$ & $ci_i^{t+1} = $NOT $En_i$\\
$CiA_i$ & $CiA_i^{t+1} = (PtcHh_i$ AND NOT $Ptc_i)$ OR $ci_i$\\
$CiR_i$ & $CiR_i^{t+1} = CiA_i$ AND (NOT $PtcHh_i$ AND $Ptc_i$)\\
$en_i$ & $en_i^{t+1} = (CiA_i$ AND NOT $CiR_i$) OR $(En_i$ AND $Dl_i)$\\
$En_i$ & $En_i^{t+1} = en_i$\\
$eya_i$ & $eya_i^{t+1} = Eya_i$ OR ($CiA_i$ AND NOT $Hth_i$)\\
$Eya_i$ & $Eya_i^{t+1} = eya_i$\\
$hth_i$ & $hth_i^{t+1} = Wg_i$ AND NOT $Eya_i$\\
$Hth_i$ & $Hth_i^{t+1} = hth_i$\\
$Wg_i$ & $Wg_i = 1$ for all $t$\\
$Dl_i$ & $Dl_i= \left\{
	       \begin{array}{ll}
		 1     & \textmd{if } i \in \textmd{ IOC domain }  \\
		 0      & \textmd{otherwise} \\
	       \end{array}
	     \right.\textmd{ for all }t$\vspace{0.5cm}\\

\emph{Model $\#2$}&\\
$Hh_i$ & $Hh_i = 1$ for all $t$\\
$PtcHh_i$ & $PtcHh_i^{t+1} = (Hh_i$ OR $CiA_i)$ AND NOT $En_i$\\
$CiA_i$ & $CiA_i^{t+1} = PtcHh_i$ AND NOT $En_i$\\
$En_i$ & $En_i^{t+1} = CiA_i$ OR $(En_i$ AND $Dl_i)$\\
$RD_i$ & $RD_i^{t+1} = (CiA_i$ OR $RD_i)$ AND NOT $Hth_i$\\
$Hth_i$ & $Hth_i^{t+1} =$ NOT $RD_i$\\
$Dl_i$ & $Dl_i= \left\{
	       \begin{array}{ll}
		 1     & \textmd{if } i \in \textmd{ IOC domain }  \\
		 0      & \textmd{otherwise} \\
	       \end{array}
	     \right.\textmd{ for all }t$\vspace{0.5cm}\\

\emph{Model $\#3$}&\\
$Hh_i$ & $Hh_i = 1$ for all $t$\\
$PtcHh_i$ & $PtcHh_i^{t+1} = (Hh_i$ OR $CiA_i)$ AND NOT $En_i$\\
$CiA_i$ & $CiA_i^{t+1} = PtcHh_i$ AND NOT $En_i$\\
$En_i$ & $En_i^{t+1} = CiA_i$ OR $(En_i$ AND $Dl_i)$\\
$Dl_i$ & $Dl_i= \left\{
	       \begin{array}{ll}
		 1     & \textmd{if } i \in \textmd{ IOC domain }  \\
		 0      & \textmd{otherwise} \\
	       \end{array}
	     \right.\textmd{ for all }t$\vspace{0.5cm}\\
\noalign{\smallskip}

\emph{Model $\#4$}&\\
$Hh_i$ & $Hh_i = 1$ for all $t$\\
$CiA_i$ & $CiA_i^{t+1} = Hh$ AND NOT $En_i$\\
$En_i$ & $En_i^{t+1} = CiA_i$ OR $(En_i$ AND $Dl_i)$\\
$Dl_i$ & $Dl_i= \left\{
	       \begin{array}{ll}
		 1     & \textmd{if } i \in \textmd{ IOC domain }  \\
		 0      & \textmd{otherwise} \\
	       \end{array}
	     \right.\textmd{ for all }t$\vspace{0.5cm}\\
\noalign{\smallskip}\hline
\end{tabular}
\label{tab:boolean}
{\\}\vspace*{-2pt}
\end{table}

\begin{appendix}
\newpage
\section{Analysis of the cooperation reactions in CiA and En} \label{sec:coop}

Equations (\ref{eq:CiA}) and (\ref{eq:En}) show the temporal evolution of CiA and En concentrations respectively. 

\begin{equation}\label{eq:CiA}
\frac{\partial CiA}{\partial \tau }=\frac{\frac{\gamma_{PtcHh-CiA}}{1+\alpha_{En-CiA}En}\alpha_{PtcHh-CiA}PtcHh^{n_{CiA}}}{k_{m_{CiA}}^{n_{CiA}}+\alpha_{PtcHh-CiA}PtcHh^{n_{CiA}}}-\beta_{CiA}CiA
\end{equation}

\begin{equation}\label{eq:En}
\frac{\partial En}{\partial \tau }=\frac{\gamma_{CiA-En}CiA^{n_{En}}}{k_{m_{En}}^{n_{En}}+CiA^{n_{En}}}+\alpha_{En-En}EnDl(\zeta )-\beta_{En}En
\end{equation}

\begin{figure}
\begin{center}
\includegraphics[width=\textwidth]{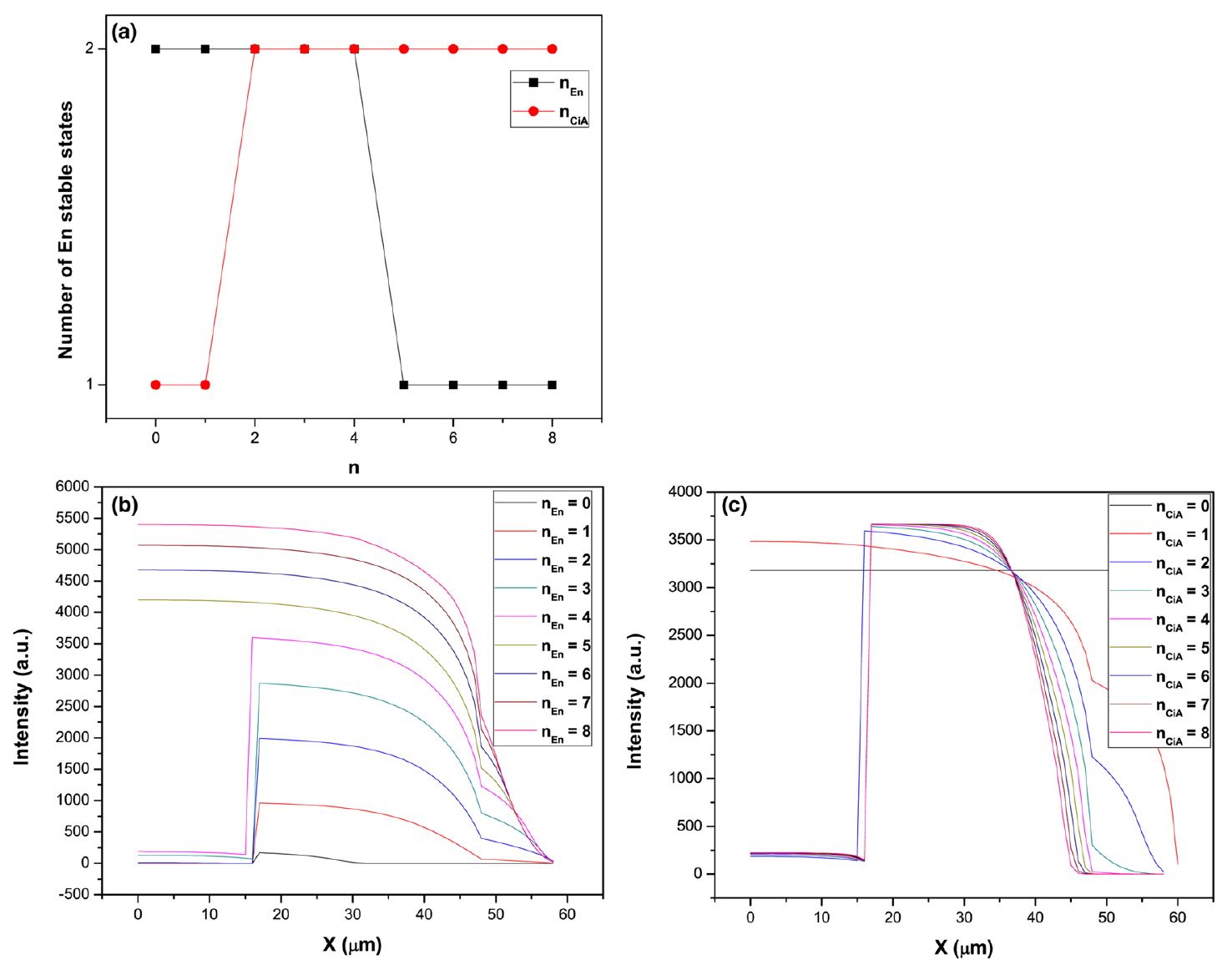}
\end{center}
\caption{(a) States plot showing the changes in En behavior for different values of the Hill coefficients $n_{En}$ and $n_{CiA}$ in equations (\ref{eq:CiA}) and (\ref{eq:En}). The vertical axis is the number of stable states that En can reach as a function of the Hill coefficient ($n$ is either $n_{En}$ as $n_{CiA}$). Black line represents the number of stable states that En reaches when modifying $n_{En}$. When $n_{En} \leq 4$, En shows just two states while for higher values, En just shows one stable state without transition to its self-regulation. Red line represents the number of stable nodes that En reaches when modifying $n_{CiA}$. In this case, En shows the transition to a second state from $n_{CiA} = 2$ on. The combination of the two Hill coefficients valid range gives a system correct behavior for $n_{En} \in [0,4]$ and $n_{CiA} \geq 2$ (b) RD profiles for different values of $n_{En}$. For $n_{En}$ values higher than 4 ($n_{En} > 4$) En is not able to self-regulate and the RD expression patterns covers the whole ocellar complex at high level. Lower $n_{En}$ values show a correct RD profile lowering its intensity while decreasing $n_{En}$. (c) RD profiles for different values of $n_{CiA}$. Only for $n_{CiA} \geq 2$ the RD profile fulfill the correct pattern. As the $n_{CiA}$ value increases the RD pattern sharpens}
\label{fig:4}
\end{figure}

In both equations, the production rate term is written as a Hill function, typically represented as:

\begin{equation}\label{eq:Hill}
 F(x)=C\frac{x^n}{k_m^n+x^n},
\end{equation}

where $k_m$ is the $x$ concentration producing half $F$; $C$ is the maximum value for $F$ and $n$ is the Hill coefficient which describes the cooperativity or affinity to continue the reaction defined with the Hill equation once started.
It is known that $n < 1$ means negative cooperation (the reaction affinity decreases), $n = 1$ means non-cooperative reaction and $n > 1$ means positive cooperation (increasing the reaction affinity).
The cooperation study was done by modifying the Hill coefficients value and observing the RD output profile, while maintaining unchanged the rest of the parameters.
RD expression pattern is only able to correctly define the IOC and OC regions when both $n_{CiA}$ and $n_{En} \in [2,4]$ (Fig. \ref{fig:4}a). This is, just in the defined Hill coefficients intervals En is able to achieve its self-regulation state transiting from one to two stable states.
When increasing $n_{En}$ over 4, the RD expression profile increases its intensity extending across the entire ocellar region (Fig. \ref{fig:4}b). Giving values for $n_{CiA} > 2$, the RD ocellus profile sharpens acquiring a more similar to square shape (Fig. \ref{fig:4}c).
Hereinafter, the Hill coefficient parameters values will be $n_{En} = 4$ and $n_{CiA} = 2$.

\section{Stability analysis for non-linear contribution equations}\label{sec:stability}

The aim of this stability analysis is double. In the first place this study can give a valid range of values for $k_m$ parameters. And secondly, this analysis will show the system behavior before and after the beginning of the En self-regulation, from a stability point of view.
There are two differential equations in which a formulation of the Hill's equation form has been used: CiA's (equation \ref{eq:CiA}) and En's (equation \ref{eq:En}).

\subsection{Stability analysis for CiA's equation} \label{sec:stabCiA}

In order to achieve a stability analysis on CiA behavior we performed a simplification on equation \ref{eq:CiA}.

\begin{equation}\label{eq:simpl_CiA}
\frac{\partial CiA}{\partial \tau }=\frac{\gamma_{PtcHh-CiA}\alpha_{PtcHh-CiA}PtcHh^{n_{CiA}}}{k_{m_{CiA}}^{n_{CiA}}+\alpha_{PtcHh-CiA}PtcHh^{n_{CiA}}}-\beta_{CiA}CiA
\end{equation} 

This equation (\ref{eq:simpl_CiA}) represents the behavior of CiA before including En into the system. This way we can also analyze the behavior of the system when introducing this repressor. It is also possible to differentiate between a linear and a non-linear repression addition.
Due to the high non-linearity of the introduced equations system, we are proposing to consider PtcHh as a direct contributor to the CiA production, including the effect of both variables into just one $x$, that will represent the full output of the Hh signaling pathway. This way, equation (\ref{eq:simpl_CiA}) can be reduced to 

\begin{equation}\label{eq:x_CiA}
\frac{\partial x}{\partial \tau }=\frac{\gamma_{PtcHh-CiA}x^{n_{CiA}}}{k_{m_{CiA}}^{n_{CiA}}+x^{n_{CiA}}}-\beta_{CiA}x\textmd{ where }x\in \left[0,13146.4\right]
\end{equation}

\begin{figure}
\begin{center}
\includegraphics[width=\textwidth]{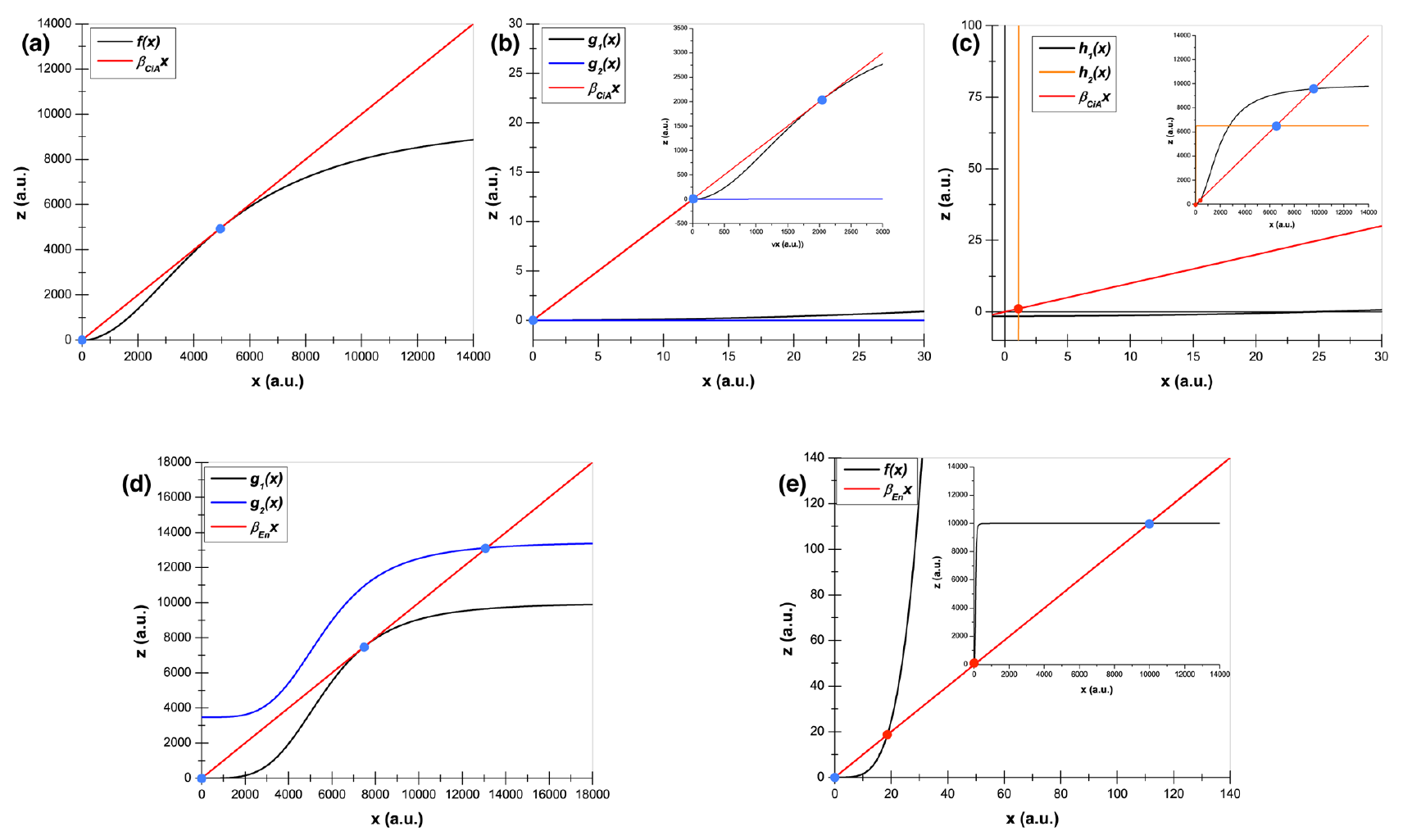}
\end{center}
\caption{Stability analysis for non-linear contribution equations. (a) This graph represents the production rate (black line) and the degradation rate (red line) of $x$ (vertical axis: $z = f(x)$, $z = \beta_{CiA}x$ in eq. \ref{eq:f_Cia}) as a function of the concentration of $x$ (horizontal axis), when En is not considered as a CiA repressor. The points at which the production rate is equal to the degradation rate are considered steady states. This configuration shows two different steady states (two stable nodes as blue dots). (b) Representation of the system steady states when En is repressing CiA in a non-linear way. Black line shows the $x$ production rate when En is in its lower chosen value (En = 1.51, before beginning its positive self-regulation). Blue line (very close to horizontal axis) represents the $x$ production rate when En has reached its higher chosen value (En = 3502.49, after En self-regulation response). Red line represents the $x$ degradation rate. In the first case the system maintains its bi-stability and in the second case the system loses the bi-stability maintaining just one stable node at $x = 0$. (c) Representation of the system steady states when En is repressing CiA in a linear way. Black line represents the $x$ production rate when En remains in its lower chosen value. Orange line represents the production rate when considering En on its higher chosen value. Red line represents the $x$ degradation. The system shows two steady states on each configuration, one stable at high $x$ value and one unstable at low $x$ value (red dot). If the system stays at the unstable steady state, small perturbations will be amplified letting the system reach the stable node. The linear repression makes the system lose its bi-stability and does not allow any stable state at low $x$ value to consider closing the Hh signaling pathway. (d) Graphical determination of the first stable steady state for En evolution (equation \ref{eq:f_En}) showing the intersection between the production rate (black line) and the degradation rate (red line). The blue line represents the production rate when En self-regulates (equation \ref{eq:g_En}). In this case, the system loses its bi-stability remaining just one stable node at high En concentration. (e) Representation of the En production rates (black line) considering the chosen value for $k_{m_{En}}=89.56$, and the degradation rate (red line). The system shows a bi-stable state with null and high $x$ levels, also considering one unstable state at low $x$ level}
\label{fig:5}
\end{figure}

The $x$ value range was taken from the numerical simulations and it is built from the union of the PtcHh and CiA value intervals.
This equation can also be expressed as:

\begin{equation}\label{eq:f_Cia}
 \frac{\partial x}{\partial \tau} = f(x)-\beta_{CiA}x
\end{equation}

where $f(x)$ is the production rate and $\beta_{CiA}x$  is the degradation rate.
Fig. \ref{fig:5}a shows a bi-stable switch where the intersection of the two tendencies exhibit two steady states (production rate is equal to the degradation rate); these steady states are stable points (in blue). So, the Hh-signaling pathway can either be closed (low CiA concentration unable to regulate RD) or open (high CiA concentration that up-regulates RD).

In this stability analysis, it was used the same equation but adding En in two possible ways: as a non-linear repression (equation \ref{eq:CiA}) and as a linear repression (equation \ref{eq:CiA_linEn}):

\begin{equation}\label{eq:CiA_linEn}
\frac{\partial CiA}{\partial \tau }=\frac{\gamma_{PtcHh-CiA}\alpha_{PtcHh-CiA}PtcHh^{n_{CiA}}}{k_{m_{CiA}}^{n_{CiA}}+\alpha_{PtcHh-CiA}PtcHh^{n_{CiA}}}-\beta_{CiA}CiA-\alpha_{En-CiA}En
\end{equation} 

Repeating the same procedure to these two equations:

\begin{equation}\label{eq:g_CiA}
\frac{\partial CiA}{\partial \tau }=\frac{\frac{\gamma_{PtcHh-CiA}}{1+\alpha_{En-CiA}En}x^{n_{CiA}}}{k_{m_{CiA}}^{n_{CiA}}+x^{n_{CiA}}}-\beta_{CiA}x=g(x)-\beta_{CiA}x=z(x)
\end{equation}

With expression \ref{eq:g_CiA} for the non-linear repression form, and

\begin{equation}\label{eq:h_CiA}
\frac{\partial CiA}{\partial \tau }=\frac{\gamma_{PtcHh-CiA}x^{n_{CiA}}}{k_{m_{CiA}}^{n_{CiA}}+x^{n_{CiA}}}-\beta_{CiA}CiA-\alpha_{En-CiA}En = h(x)-\beta_{CiA}x=z(x)
\end{equation} 

being the expression \ref{eq:h_CiA} for the linear repression form.\\

Two different values were chosen for En concentration depending on its state: without self-regulation (1.51) or with self-regulation (3502.49). These values were taken from numerical simulations being closed to a steady state. 
In order to obtain a value for $k_{m_{CiA}}$ that ensures at least one stable steady state, the second derivative of $z(x)$ must be negative in a $\left\lbrace x,k_{m_{CiA}} \right\rbrace$ valid value pair. This value pair is the tangent point between $g(x)$ and $\beta_{CiA}x$.
The first $z(x)$ derivative from equation \ref{eq:g_CiA} is:

\begin{equation}\label{eq:zder1}
 \frac{dz}{dx}=-\frac{\frac{2\gamma_{PtcHh-CiA}}{1+\alpha_{En-CiA}En}x^3}{\left ( k_{m_{CiA}}^{2}+x^2 \right )^2}+\frac{\frac{2\gamma_{PtcHh-CiA}}{1+\alpha_{En-CiA}En}x}{ k_{m_{CiA}}^{2}+x^2}-\beta_{CiA}
\end{equation}

From equations (\ref{eq:g_CiA}) and (\ref{eq:zder1}) is obtained, as the only real and positive solution, that the $x$ production and degradation rate are tangent at $x = k_{m_{CiA}}= 2000$. For these values the system ensures a positive stable node and a valid CiA behavior. This is, CiA concentration is high when En is not self-regulated (black curve in Fig.\ref{fig:5}b), and CiA concentration is very low or null for high concentration of En (when En is self-regulated; blue curve in Fig. \ref{fig:5}b). So the closure of the Hh-signaling pathway can be achieved when En represses CiA in a non-linear way.
The second $z(x)$ derivative is negative at the calculated $x$ and  $k_{m_{CiA}}$ value, ensuring its convexity.

\begin{equation}\label{eq:zder2}
 \frac{d^2z}{dx^2}=\frac{\frac{8\gamma_{PtcHh-CiA}}{1+\alpha_{En-CiA}En}x^{4}}{\left ( k_{m_{CiA}}^{2}+x^2 \right )^3}-\frac{\frac{10\gamma_{PtcHh-CiA}}{1+\alpha_{En-CiA}En}x^2}{\left ( k_{m_{CiA}}^{2}+x^2 \right )^2}+\frac{\frac{2\gamma_{PtcHh-CiA}}{1+\alpha_{En-CiA}En}}{ k_{m_{CiA}}^{2}+x^2 }
\end{equation}

For $k_{m_{CiA}} = 2000$, $\frac{d^2z}{dx^2}\left(x=2000\right)=-2\cdot 10^{-5}$.\\

When evaluating the En repression action to CiA in a linear way (eq. \ref{eq:h_CiA}), we find the same behavior independently of En activation. Either En being at low or high concentration it is shown one unstable steady state at low $x$ values and a stable steady state at high $x$ levels (Fig. \ref{fig:5}c). With this linear repression the Hh-signaling pathway can remain open with a high concentration of CiA in the cells where En is self-regulated. Moreover, as the steady state found at low $x$ values is unstable, perturbations applied to the system can lead its behavior to the stable, and high CiA concentration node. In other words, if the system is able to close the Hh-signaling pathway due to a linear En repression, it can be spontaneously open again if the system is perturbed. This behavior does not correspond to the CiA experimentally observed in \cite{AguilarHidalgo:2013bw}. It is known that CiA reaches a high concentration positive stable node when En is not self-maintained and that it is overridden when En self-regulation is activated.
In the case of non-linear repression (eq. \ref{eq:g_CiA}) $x$ shows a bi-stable state when En is not self-regulated. One stable node at $x = 0$ and a second one at $x = 2000$. When En self-regulation starts $x$ loses its bi-stability maintaining just the stable steady state at null $x$ levels (Fig. \ref{fig:5}b). This behavior fits with the experimental one. The only possible stable node of CiA when En self-maintains is the closure of Hh-signaling pathway, this is, CiA concentration falls down. 

\subsection{Stability analysis for En's equation} \label{sec:stabEn}

A similar analysis can be done to the differential equation that drives the behavior of En (equation \ref{eq:En}). If En is not self-regulated ($Dl = 0$), this equation can be expressed as:

\begin{equation}\label{eq:f_En}
\frac{\partial x}{\partial \tau }=\frac{\gamma_{CiA-En}x^{n_{En}}}{k_{m_{En}}^{n_{En}}+x^{n_{En}}}-\beta_{En}x=f(x)-\beta_{En}x=z(x)
\end{equation}

The z(x) first derivative of equation \ref{eq:f_En} is:

\begin{equation}\label{eq:zEnder1}
 \frac{dz}{dx}=-\frac{4\gamma_{CiA-En}x^7}{\left ( k_{m_{Em}}^{4}+x^4 \right )^2}+\frac{4\gamma_{CiA-En}x^3}{k_{m_{En}}^{4}+x^4}-\beta_{En}
\end{equation}

The tangent point in this case is $x=7500$ and $k_{m_{En}}=5698.77$.

The second derivative (eq. \ref{eq:zEnder2}) shows that the tangent point is a stable one as $\frac{d^2 z}{d x^2}=-4\cdot 10^{-4}$ as corresponds to a maximum.

\begin{equation}\label{eq:zEnder2}
 \frac{d^2z}{dx^2}=\frac{32\gamma_{CiA-En}x^{10}}{\left ( k_{m_{En}}^{4}+x^4 \right )^3}-\frac{44\gamma_{CiA-En}x^6}{\left ( k_{m_{En}}^{4}+x^4 \right )^2}+\frac{12\gamma_{CiA-En}x^2}{k_{m_{En}}^{4}+x^4}
\end{equation}

This way the system shows again a bi-stable state with a steady state at $x=0$ and another at $x=7500$ (Fig. \ref{fig:5}d - black curve). Every value for the parameter $k_{m_{En}}\leq 5698.77$ is valid to ensure this bi-stability.
The chosen value for this parameter is $k_{m_{En}}= 89.56$, this way the RD pattern is closer to the experimental one.  By using this value, the system maintains its bi-stability with an intermediate unstable steady state at a very low $x$ value (Fig. \ref{fig:5}e). Once En self-regulates ($Dl = 1$, eq. \ref{eq:g_En}) the system loses its bi-stability maintaining just one stable steady state at high $x$ value (Fig. \ref{fig:5}d - blue curve). In other words, once En self-regulates, its concentration can only be high, and, as commented, high enough to make CiA fall down triggering the closure of the Hh-signaling pathway.

\begin{equation}\label{eq:g_En}
\frac{\partial x}{\partial \tau }=\frac{\gamma_{CiA-En}x^{n_{En}}}{k_{m_{En}}^{n_{En}}+x^{n_{En}}}+\alpha_{En-En}En-\beta_{En}x=f(x)-\beta_{En}x=z(x)
\end{equation}

\section{The ocellar regulatory network is robust facing noise}

Complex networks usually include network motifs that may infer robustness to the system capable of withstanding the presence of noise and maintain its original behavior \cite{Nahmad:2009cp}. This well known behavior raises the question of whether the ocelli GRN is capable to withstand noisy perturbations and still maintain its qualitative behavior. In order to \emph{in silico} demonstrate that the ocelli GRN is robust against noise we have implemented some computational experiments in which a random component was inserted in the system to simulate white noise on the Hh signaling-pathway. These experiments applied a white noise ($\nu$) (uniform random distribution centered at one) to CiA production term (eq. \ref{eq:CiA-noise}). The standard deviation ($\sigma$) of the distribution was varied from 5\% to 20\% by 5\%. The highest noise proportion should be large enough to be considered superior to biological noise. 

\begin{equation}\label{eq:CiA-noise}
\frac{\partial CiA}{\partial \tau }=\frac{\frac{\gamma_{PtcHh-CiA}}{1+\alpha_{En-CiA}En}\alpha_{PtcHh-CiA}PtcHh^{n_{CiA}}}{k_{m_{CiA}}^{n_{CiA}}+\alpha_{PtcHh-CiA}PtcHh^{n_{CiA}}}\nu(\sigma)-\beta_{CiA}CiA
\end{equation}

\begin{figure}
\begin{center}
\includegraphics[width=\textwidth]{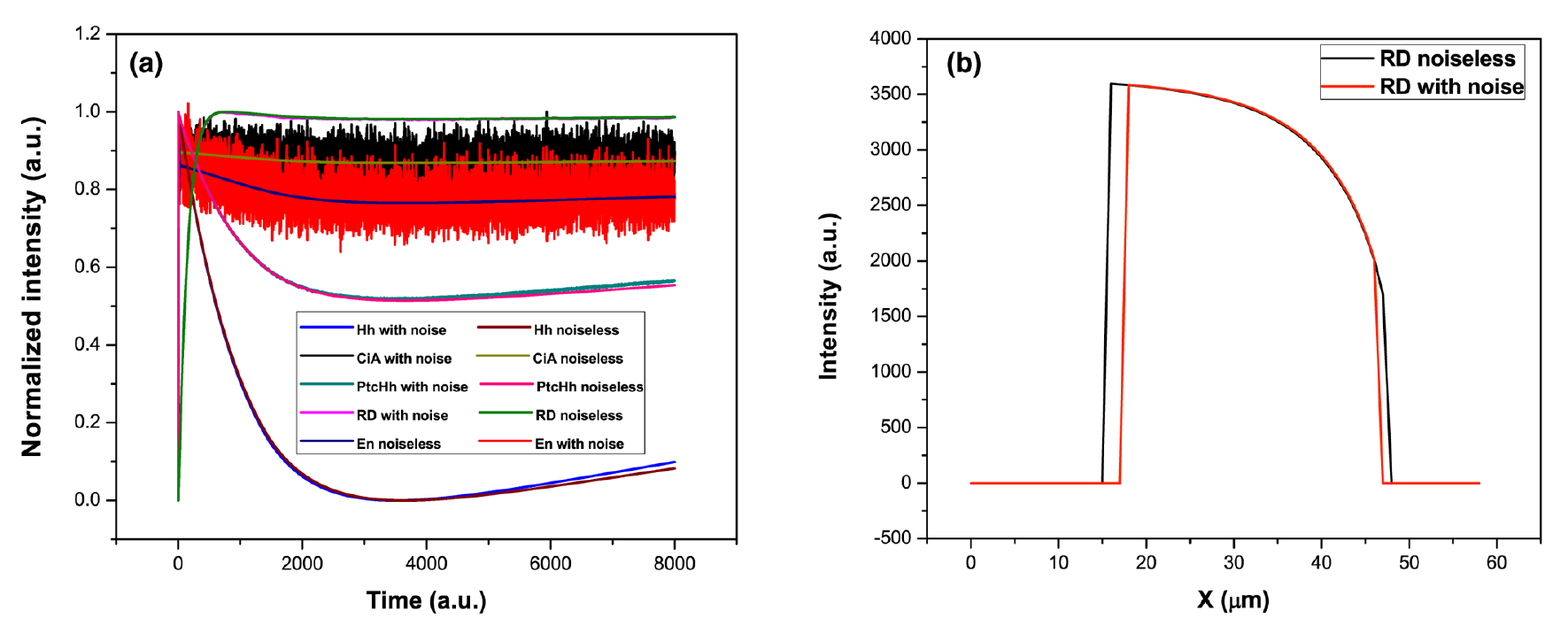}
\end{center}
\caption{(a) Time series of the systems elements showing the difference of each signal when applying a white noise (flat random distribution) to the end of the Hh-signaling pathway (CiA node) with an amplitude of the 20\% of the original production term value. This plot shows that the system is robust against noise as the signals remain qualitatively, and nearly quantitatively, the same after applying the white noise distribution. (b) RD profile showing one single ocellus pattern applying white noise and without applying white noise. This plot shows that the pattern remains qualitatively the same and that the ocellus is slightly smaller when applying the white noise distribution}
\label{fig:6}
\end{figure}

By adding this noisy signal to CiA, it is intended to simulate conditions in which variations in the concentration of CiA, as the readout of the Hh-signaling pathway, may infer differences in the real ocellar patterning due to changes in the environmental conditions during the developmental process. The obtained results show that, even in the most dramatic situation, the behavior of the system is qualitatively the same as in the case without noise. Only at the maximum noise proportion, the system shows small quantitative differences in PtcHh and Hh concentration, but never qualitative (Fig. \ref{fig:6}a – only shown for 20\% of noise). For this maximum noise proportion it is also observed a quantitative difference in RD concentration, showing a narrower profile when considering CiA noisy signal (Fig. \ref{fig:6}b). It was also tested if the presence of Hth alters the noise effect in the RD profile. The result was that no changes appeared in the noise effect due to Hth presence, so Hth does not infer any buffer effect to RD signal. 

\section{Analysis of the Hh gradient steady state}

It is known that \emph{hh} expression is limited to the interocellar region named as \emph{hh}-expressing zone. This region has a width of $\omega=33\mu m$ and it is centered in $x=0$ in this model. It has been described that in this region $\delta(x) = 1$, so $\omega$ is the size of the morphogen source.
Once the morphogen gradient reaches its steady state ($\tau = \tau_{std}$) we have:

\begin{equation}\label{eq:Hh_std}
 \frac{\partial Hh}{\partial \tau }\Bigg|_{\tau=\tau_{std}}=0
\end{equation}

so,

\begin{equation}\label{eq:Hh_eq_std}
 D\frac{\partial^2 Hh}{\partial x^2 }+\delta(x)\alpha_{hh}-\alpha_{PtcHh-Hh}PtcHh-\beta_{Hh}Hh=0
\end{equation}

In the steady state, we can consider that PtcHh concentration is also constant in time, so this can be treated as a constant value for this calculation. To solve this diffusion equation with degradation and a source of width $\omega$ centered in $x=0$ it is needed to formalize the contour conditions. In the first place, it will be considered vanishing concentrations for $x=\pm L$, being $2L$ the whole one-dimensional field of cells that we are working on. We defined $2L=117\mu m$. This assumption make sense if the characteristic length of the gradient $\lambda$ is much smaller than $L$. Secondly, it will be considered that both the morphogen concentration, $Hh(x)$, and its flux, $j=-D\frac{\partial Hh}{\partial x}+\beta_{Hh}Hh$ must be continuous at the boundaries of the morphogen source ($x=\pm \omega/2$). Taken into account these considerations the solution for equation \ref{eq:Hh_eq_std} is:

\begin{equation} \label{eq:full_std_sol}
Hh(x) = \left\{
	       \begin{array}{ll}
		 Hh_b e^{\left(x+\frac{\omega}{2}\right)\lambda^{-1}}-S_b      & \textmd{if\ } x \leq-\frac{\omega}{2} \\
		 \frac{\alpha_{hh}}{2\beta_{Hh}}\left(2-e^{\left(x-\frac{\omega}{2}\right)\lambda^{-1}}-e^{\left(-x-\frac{\omega}{2}\right)\lambda^{-1}}\right)-S_b      & \textmd{if\ } x -\frac{\omega}{2}\leq x\leq \frac{\omega}{2} \\
		 Hh_b e^{\left(-x+\frac{\omega}{2}\right)\lambda^{-1}}-S_b      & \textmd{if\ } x \geq\frac{\omega}{2} \\
	       \end{array}
	     \right.
\end{equation}

With $Hh_b=\frac{\alpha_{hh}}{2\beta_{Hh}}\left(1-e^{-\frac{\omega}{\lambda}}\right)$ and $S_b = \frac{\alpha_{PtcHh-Hh}PtcHh}{\beta_{Hh}}$, being $Hh_b-S_b$ the Hh concentration at the boundaries of the source. $\lambda=\sqrt{D/\beta_{Hh}}$ is also known as decay length. This last corresponds to the distance at which the morphogen concentration decays by a factor of $1/e$. It can be calculated from the parameters used in the simulation (see Table \ref{tab:params}) that the value for $\lambda=0.7\mu m$, so $\lambda<<L$.

\subsection{Hh morphogen gradient fits a Gaussian distribution}

It has been shown that the analytical solution for the Hh morphogen equation on its stationary state can be written as a combination of exponential functions (equation \ref{eq:full_std_sol}). From an experimental perspective, it is known that morphogen gradients profiles, which spatial distribution is similar to a 1D profile with a non-punctual source, can follow a Gaussian distribution or a sum of Gaussian distributions. An example of this is the Dpp morphogen gradient in the Drosophila wing disc \cite{Kicheva:2007bha} (see Appendix D). The same way, we can treat the numerical output as an experimental profile and fit it to a Gaussian distribution. This distribution expression is the following:

\begin{equation}\label{eq:gauss}
 Hh(x) = A_0+\frac{A}{\sigma\sqrt{2\pi}}e^{-\frac{1}{2}\left(\frac{x-x_c}{\sigma}\right)^2}
\end{equation}

In Fig. 4a is shown the comparison between the numerical stationary solution of the Hh morphogen gradient and the Gaussian distribution fitting with $A_0=-810\pm50$, $A=6.620\cdot10^6\pm6\cdot10^3$, $\sigma=23.68\pm0.15$ and $x_c=0$. The errors in these parameters come from the fitting calculation with a R-squared value of $R = 0.99863$.

 \begin{table}[hbt]
 \caption{Parameter values}
 \begin{tabular*}{\textwidth}{@{}lll@{}}
 \hline
 Parameter & Description &Value\\
 \hline
 $D$ & Hh diffusion constant & $0.5\mu m^2 s^{-1}$, \cite{Nahmad:2009cp}\\
$\alpha_{hh}$ &Hh transcription rate & $10\mu M s^{-1}$ \\
$\alpha_{Hh-PtcHh}$ & Association rate for PtcHh complex formation & $1s^{-1}$ \\
$\alpha_{PtcHh-Hh}$ & Ptc efficiency to capture Hh & $5\cdot 10^{-4} s^{-1}$ \\
$\alpha_{PtcHh-CiA}$ & PtcHh activating CiA rate & $1$ \\
$\alpha_{CiA-PtcHh}$ & CiA activating PtcHh rate & $0.1 s^{-1}$\\
$\alpha_{CiA-RD}$ & CiA activating RD rate  &  $1s^{-1}$  \\
$\alpha_{En-PtcHh}$ & En repressing PtcHh rate  &  $1s^{-1}$   \\
$\alpha_{En-CiA}$ & En repressing CiA rate  &  $0.99$ \\
$\alpha_{En-En}$ & En translation rate  &  $1s^{-1}$ \\
$\alpha_{RD-Hth}$ & RD repressing Hth rate  &  $0.001s^{-1}$  \\
$\alpha_{Hth-RD}$ & Hth repressing RD rate  &  $0.1s^{-1}$ \\
$\alpha_{RD-RD}$ &  RD auto-regulation rate  & $0.99s^{-1}$  \\
$\alpha_{Wg-Hth}$ &  Wg activating Hth rate   & $4.7\mu Ms^{-1}$  \\
$Dl$ & En self-activation switch  &  1-ON, 0-OFF  \\
$\zeta_{En}$ &  En threshold for Dl activation  & $256\mu M$  \\
$\zeta_{t}$ &  Time interval with En over $\zeta_{En}$ for Dl activation  & 1400s  \\
$\gamma_{PtcHh-CiA}$ & Maximum CiA production rate  &  $10000\mu Ms^{-1}$   \\
$\gamma_{CiA-En}$ & Maximum En activation by CiA  &  $10000\mu Ms^{-1}$   \\
$\beta_{Hh}$ & Hh degradation rate  &  $1s^{-1}$   \\
$\beta_{PtcHh}$ & PtcHh degradation rate  &  $1s^{-1}$   \\
$\beta_{CiA}$ & CiA degradation rate  &  $1s^{-1}$   \\
$\beta_{En}$ & En degradation rate  &  $1s^{-1}$   \\
$\beta_{RD}$ & RD degradation rate  & $1s^{-1}$   \\
$\beta_{Hth}$ & Hth degradation rate  & $0.03s^{-1}$   \\
$k_{m_{CiA}}$ & CiA half-maximal activation concentration  & $2000\mu M$ (sec. \ref{sec:stabCiA})  \\
$k_{m_{En}}$ & En half-maximal transcriptional activation  & $89.56\mu M$ (sec. \ref{sec:stabEn})    \\
$n_{CiA}$ & Hill coefficient for CiA transcription  & 2 (sec.\ref{sec:coop})   \\
$n_{En}$ & Hill coefficient for En transcription  & 4  (sec. \ref{sec:coop}) \\

\noalign{\smallskip}\hline
\end{tabular*}
\label{tab:params}    

\end{table}

\begin{table}[hbt]
\caption{Variables initial conditions}
\begin{tabular}{@{}lll@{}}
\hline\noalign{\smallskip}
Variable & Description & Initial condition  \\
\noalign{\smallskip}\hline\noalign{\smallskip}
Hh & Hh concentration & 0 \\
PtcHh & PtcHh complex concentration & $100\mu M$ \\
CiA & CiA concentration & 0 \\
En & En concentration & $1\mu M$ \\
RD & RD concentration & $1\mu M$ \\
Hth & Hth concentration & $156.67\mu M^a$ \\
\noalign{\smallskip}\hline
\end{tabular}
\label{tab:ci}
{\\$^a$ Initial Hth concentration corresponds to its stationary concentration in the absence of RD repression ($[Hth]=\alpha_{Wg-Hth}/\beta_{Hth}$, see table \ref{tab:params}).}\vspace*{-2pt}
\end{table}

\end{appendix}

%\appendix

%\section{Section in Appendix}
%\label{appendix-sec1}

%Sample text. Sample text. Sample text. Sample text. Sample text. Sample text. 
%Sample text. Sample text. Sample text. Sample text. Sample text. Sample text. 
%Sample text. 

%% References
%%   \citet{key}         ==>>  [#]


%\bibliographystyle{plain}
%\bibliographystyle{model2-names.bst}\biboptions{authoryear}
%5\bibliographystyle{plainnat}
%\bibliographystyle{elsarticle-num}%\biboptions{number}
 %\bibliographystyle{elsarticle-harv}
% \bibliographystyle{elsarticle-num-names}
% \bibliographystyle{model1a-num-names}
% \bibliographystyle{model1b-num-names}
% \bibliographystyle{model1c-num-names}
% \bibliographystyle{model2-names}
% \bibliographystyle{model3a-num-names}
% \bibliographystyle{model3-num-names}
% \bibliographystyle{model4-names}
% \bibliographystyle{model5-names}
% \bibliographystyle{model6-num-names}

%\bibliography{ocellist}

\end{document}